\documentclass[11pt]{article}

\usepackage{scienta}

\usepackage[utf8]{inputenc} % allow utf-8 input
\usepackage[T1]{fontenc}    % use 8-bit T1 fonts
\usepackage{hyperref}       % hyperlinks
\usepackage{url}            % simple URL typesetting
\usepackage{booktabs}       % professional-quality tables
\usepackage{multirow}       % span miltiple rows vertically
\usepackage{amsfonts}       % blackboard math symbols
\usepackage{nicefrac}       % compact symbols for 1/2, etc.
\usepackage{microtype}      % microtypography
\usepackage{xcolor} % for \textcolor
\usepackage{graphicx}
\usepackage{todonotes} % for \todo
\presetkeys{todonotes}{inline}{} % make todo inline by default
\usepackage{float} % for H figure placement
\usepackage{amsmath}
\usepackage{cite} % compress citations
\usepackage{adjustbox}
\usepackage{caption}
\usepackage{subcaption}
\definecolor{yellow_sl}{HTML}{FBD495}
\definecolor{orange_sl}{HTML}{FD8457}
\definecolor{blue_sl}{HTML}{234AAE}
\papertitle{EVA: Towards a universal model of the immune system}
\papersubtitle{Technical Report}

\paperauthors{Scienta Team}

\paperaffiliations{Paris, France}

\papernotes{}

\paperabstract{The effective application of foundation models to translational research in immune-mediated diseases requires multimodal patient-level representations that can capture complex phenotypes emerging from multicellular interactions. Yet most current biological foundation models focus only on single-cell resolution and are evaluated on technical metrics often disconnected from actual drug development tasks and challenges. Here, we introduce EVA, the first cross-species, multimodal foundation model of immunology and inflammation, a therapeutic area where shared pathogenic mechanisms create unique opportunities for transfer learning. EVA harmonizes transcriptomics data across species, platforms, and resolutions, and integrates histology data to produce rich, unified patient representations. We establish clear scaling laws, demonstrating that increasing model size and compute translates to improvements in both pretraining and downstream tasks performance. We introduce a comprehensive evaluation suite of 39 tasks spanning the drug development pipeline: zero-shot target efficacy and gene function prediction for discovery, cross-species or cross-diseases molecular perturbations for preclinical development, and patient stratification with treatment response prediction or disease activity prediction for clinical trials applications. We benchmark EVA against several state-of-the-art biological foundation models and baselines on these tasks, and demonstrate state-of-the-art results on each task category. Using mechanistic interpretability, we further identify biological meaningful features, revealing intertwined representations across species and technologies. We release an open version of EVA for transcriptomics to accelerate research on immune-mediated diseases.}

\begin{document}

\makeheader

\begin{center}
    \includegraphics[width=.89\textwidth]{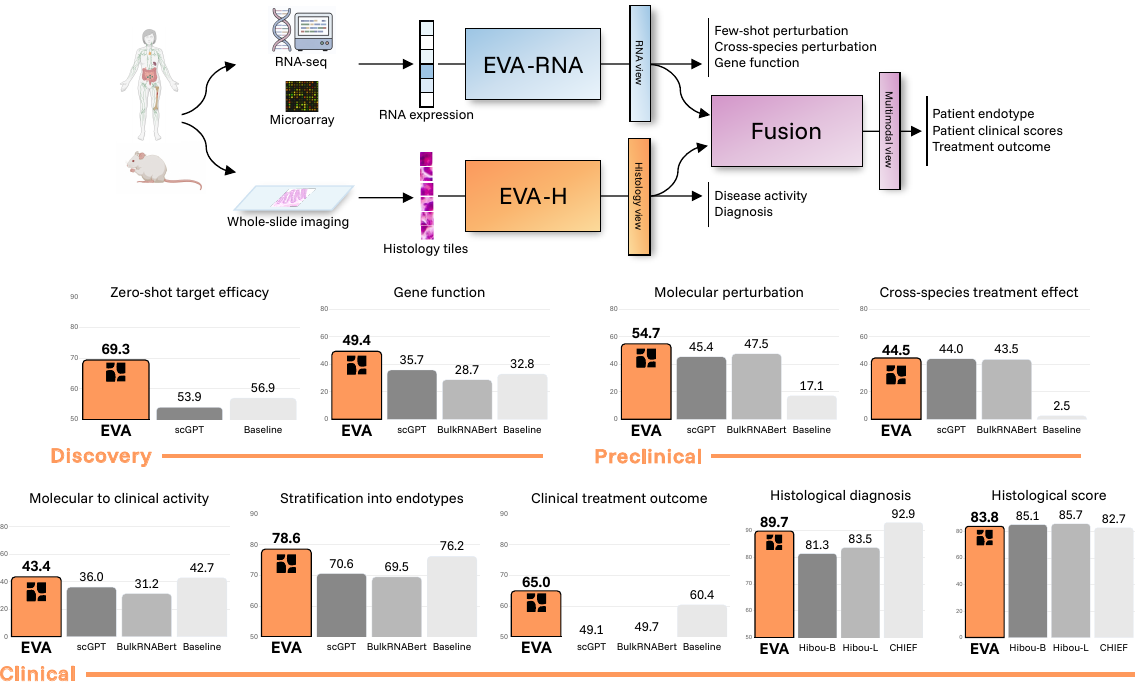}
\end{center}

\section{Introduction}

The explosion of publicly available biological data across imaging and molecular modalities, including next-generation sequencing, presents both an unprecedented opportunity and a fundamental challenge. Yet, each modality captures only a partial view of biological states, and methods to integrate these complementary perspectives remain underdeveloped. Biological foundation models have emerged as a promising paradigm for learning rich representations from large-scale data \cite{theodoris2024perspectives}, but current approaches operate predominantly within single modalities, with notable contributions in transcriptomics \cite{rosen2023universal,theodoris2023transfer,cui2024scgpt,hao2024large}, histology \cite{filiot2023scaling,nechaev2024hibou,chen2024towards}, genomics \cite{dalla2025nucleotide,zhou2023dnabert,nguyen2023hyenadna,nguyen2024sequence,avsec2025alphagenome}, and proteins \cite{lin2023evolutionary,elnaggar2021prottrans,hayes2025simulating,ferruz2022protgpt2}, leaving cross-modal integration relatively underexplored. While recent efforts have begun bridging modalities such as joint histology-transcriptomics models \cite{chen2025visual,jaume2024modeling} and multimodal protein models like ESM-3 \cite{hayes2025simulating}, systematic integration across the full spectrum of biological data types remains nascent, and the complementary insights such integration could unlock are largely untapped.

Within transcriptomics in particular, much effort has converged on high-resolution single-cell modeling (often referred to as \textit{virtual cell} \cite{bunne2024build}). Recent benchmarks reveal that these single-cell models often fail to outperform simpler baselines for relevant downstream tasks, especially in out-of-distribution scenarios \cite{kedzierska2025zero, ahlmann2025deep}, exposing a possible misalignment between the representations learned during pretraining and those required for effective transfer learning. Foundation models in other modalities face distinct challenges: histology models, despite demonstrating clear improvements over prior methods, often struggle to generalize outside of oncology, which remains the dominant data source \cite{gallagher2024going}; protein and genomics models similarly show variable transfer learning capabilities across biological contexts \cite{li2024feature,feng2025benchmarking}. Recent community efforts have started establishing standardized evaluation frameworks \cite{wu2025biology,roohani2025virtual}, yet the field still lacks meaningful benchmarks for drug discovery and translational research, comparable to ImageNet or CASP, which catalyzed breakthroughs in computer vision and protein structure prediction, respectively.

In this work, we introduce \textbf{EVA}, the first cross-species, multimodal foundation model of immunology and inflammation (I\&I), a therapeutic area characterized by cross-species conservation of disease-associated mechanisms, including cytokine signaling networks (TNF, JAK-STAT), overlapping genetic susceptibility loci, and common effector cell populations \cite{lincoln2024genetic,gonzalez2020approaching}, thereby enabling unique opportunities for transfer learning. EVA produces patient-level representations and is built around a unified transcriptomics encoder, primed with an immunology-specific histology model, and a cross-modal head trained on frozen representations from each encoder. Our contributions span model architecture and initialization, training methodology, downstream tasks alignment, evaluation and interpretability.

\begin{itemize}
    \item EVA is a 440M-parameter model (300M-parameters gene expression encoder, 85M-parameter histology encoder, 55M-parameter fusion head) that integrates human and mouse bulk RNA-seq, microarray, pseudobulked single-cell, and histology into unified sample embeddings across more than 50 tissues and conditions.
    \item We curate a comprehensive I\&I benchmark of 39 tasks spanning the drug discovery pipeline: zero-shot target efficacy and gene function predictions (discovery), cross-species, cross-conditions or cross-tissue molecular perturbations translation (preclinical), and patient stratification with treatment response prediction or molecular to clinical disease activity mapping (clinical).
    \item For EVA-RNA, our transcriptomics encoder, we establish predictable scaling behavior up to 300M parameters with no sign of plateauing and highlight that in almost all cases, pretraining validation loss improvements translate into better benchmark performance.
    \item Using sparse autoencoders with top-k activation, we identify interpretable features that reveal intertwined representations across species and technologies.  
\end{itemize}

\noindent Along with this manuscript, we release an open version of EVA-RNA to \href{https://huggingface.co/ScientaLab/eva-rna}{HuggingFace} to accelerate research in computational immunology and drug discovery.

%{\emoji{hugging-face} HuggingFace}

\begin{figure}[htbp]
    \centering
    \includegraphics[width=.92\textwidth]{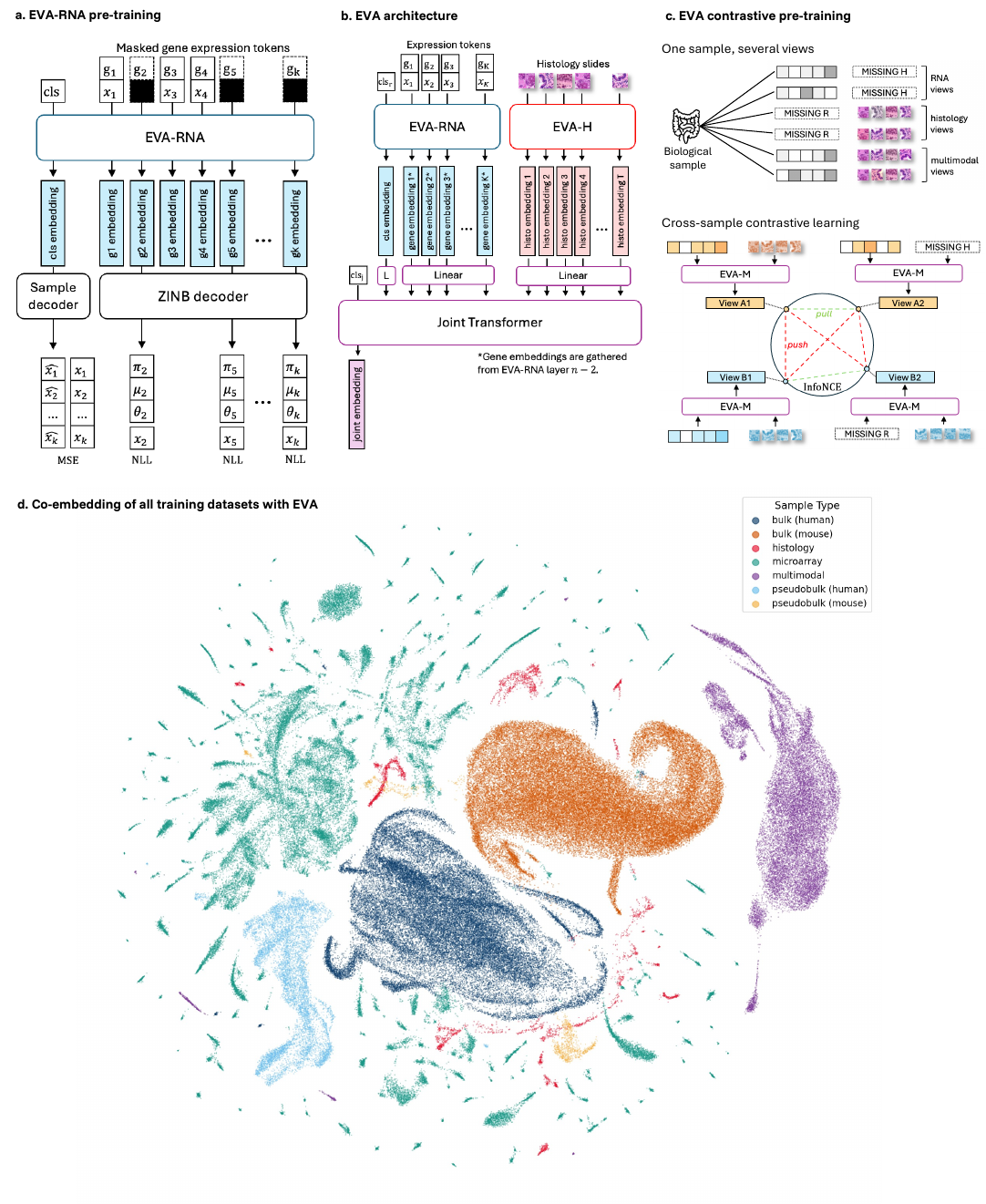}
    \caption{The EVA model architecture. (a) EVA-RNA pretraining with stochastic masked gene expression prediction and CLS token compression. (b) EVA multimodal architecture integrating gene embeddings from EVA-RNA with tile embeddings from EVA-H via a joint transformer. (c) EVA multimodal contrastive pretraining with multiple views per sample and InfoNCE objective. (d) UMAP of the training datasets embedded through EVA showing co-embedding of samples across species, technologies, and modalities. Interestingly, different modalities are embedded separately, a phenomenon observed and documented in other multimodal approaches like CLIP \cite{liang2022mind}.}
    \label{fig:global-architecture}
\end{figure}

\section{Results}

\subsection{EVA achieves state-of-the-art performance on a holistic I\&I benchmark}
We evaluated EVA on a large benchmark of 39 tasks across key steps of drug development: discovery, preclinical, and clinical areas, with their associated challenges and unique datasets. Our benchmark spans across 8 I\&I diseases involving different organs and tissues. Transcriptomics-related tasks were evaluated using the EVA-RNA encoder, and histology-related tasks leveraged EVA-H tile embeddings. We demonstrate clear improvements over both statistical baselines and existing transcriptomics foundation models, both for single-cell and bulk RNA-seq, on all task categories, as reported in Table~\ref{tab:benchmark-unimodal}. EVA is especially strong for treatment outcome prediction or endotype classification, where existing foundation models are outperformed by a simple logistic regression used as a statistical baseline. Our histology model is competitive with existing state-of-the-art models, and demonstrates strong performances in histopathological diagnosis or activity scoring (Table~\ref{tab:histo_results_aggregated}). Details of the benchmark and associated tasks are presented in Section~\ref{sec:benchmark}.

\begin{table}[htbp]
    \centering
    \caption{EVA-RNA performance on I\&I transcriptomics tasks. Bold and underline represent the best and second best models. We could not perform zero-shot target efficacy prediction with BulkRNABert as the model decoder is not publicly available.}
    \label{tab:benchmark-unimodal}
    \begin{tabular}{llcccccc}
        \toprule
        \tableheaderstyle
        Application & Task type & 7M & 60M & 300M & scGPT & BulkRNABert & Stat Baseline\\
        \midrule
        \multirow{2}{*}{Discovery} & Zero-shot target efficacy & \underline{0.655} & 0.619 & \textbf{0.693} & 0.539 & -- & 0.569 \\
         & Gene Function & 0.387& \underline{0.441}& \textbf{0.494}& 0.357& 0.287& 0.328\\
        \cmidrule(lr){1-8}
        \multirow{2}{*}{Preclinical} & Molecular perturbation & 0.511 & \underline{0.544} & \textbf{0.547} & 0.454 & 0.475 & 0.171 \\
         & Cross-Species treatment effect & 0.443 & \underline{0.444} & \textbf{0.445} & 0.439 & 0.435 & 0.025 \\
        \cmidrule(lr){1-8}
        \multirow{3}{*}{Clinical} & Molecular to clinical activity & 0.383 & \underline{0.429} & \textbf{0.434} & 0.360 & 0.312 & 0.427 \\
         & Stratification into endotypes & 0.767 & \textbf{0.799} & \underline{0.786} & 0.706 & 0.695 & 0.762 \\
         & Clinical treatment outcome & 0.572 & \underline{0.613} & \textbf{0.650} & 0.491 & 0.497 & 0.604 \\
        \bottomrule
    \end{tabular}
\end{table}

\begin{table}[htbp]
\centering
\caption{EVA-H performances on I\&I tasks. Bold and underline represent the best and second best models. Detailed results and methods can be found in Section~\ref{apx:histo-eval-method}.}
\label{tab:histo_results_aggregated}
\begin{tabular}{lccccc}
\toprule
\tableheaderstyle Application & Task type & EVA-H & Hibou-B & Hibou-L & CHIEF \\
\midrule
Clinical & Diagnosis & $\underline{0.897}$ & $0.813$ & $0.835$ & $\mathbf{0.929}$ \\
Clinical & Histological scoring & $0.838$ & $\underline{0.851}$ & $\mathbf{0.857}$ & $0.827$ \\
\bottomrule
\end{tabular}
\end{table}

\paragraph{Predicting efficacy of new targets in zero-shot settings.}

Zero-shot target efficacy prediction tasks evaluate whether pretrained representations generalize to new disease-drug combinations without task-specific fine-tuning, a scenario that mirrors the clinical reality where novel therapeutics or new indications lack historical training data.  

The task predicts whether inducing or repressing the expression of a drug's molecular target will benefit patients with a given disease. To simulate these interventions computationally, we leverage the decoder gradients of our RNA foundation model to perform \textit{in silico} gene perturbations, either target downregulation or overexpression on patient transcriptomes, following the approach proposed by Bjerregaard et al. \cite{genes16121439} (see Methods~\ref{sec:zsp}).

For evaluation purposes, we constructed a matrix spanning 28 drugs (each annotated with one or more molecular targets) across six diseases, capturing whether a given drug-disease pairing demonstrated clinical efficacy or failed to do so; entries were incomplete for a subset of combinations due to limited trial evidence (see Appendix~\ref{apx:zsp-benchmark}). Negative controls were added in the form of five additional unrelated (biologically implausible) drug targets not expected to be involved in the selected disease contexts according to field experts. 

\begin{figure}[ht!]
    \centering
    \includegraphics[width=\textwidth]{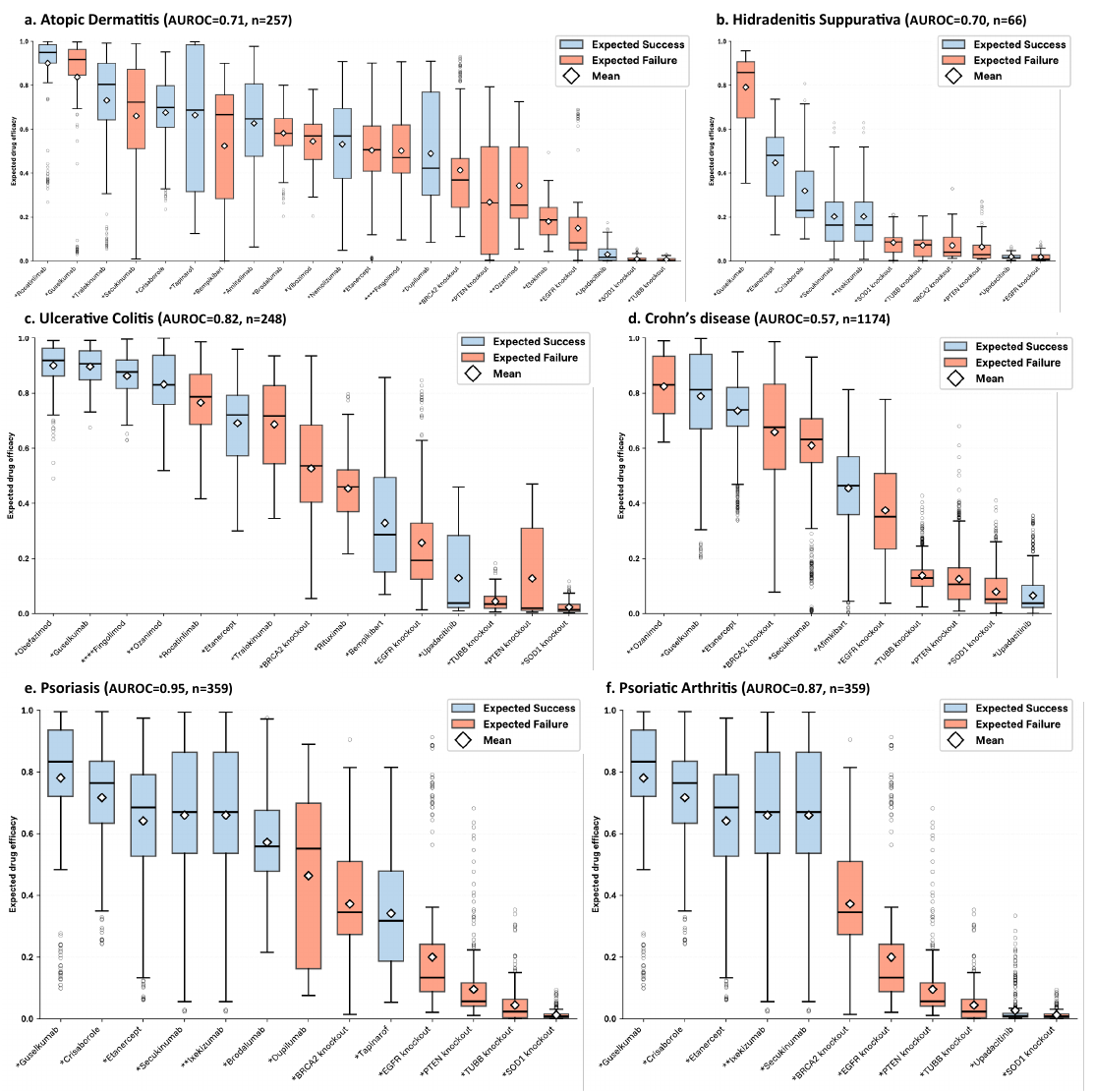}
    \caption{Zero-shot drug efficacy predictions, for each disease. The number of stars indicates the number of targets perturbed for this drug. Each box plot represents the distribution of predicted efficacy over the whole cohort. Drugs are ranked by median predicted efficacy. Blue bar plots represent drugs with confirmed positive trial results, red bar plots represent drugs with negative results or no expected efficacy. $n$ stands for the number of patients in each cohort. Detailed methodology is reported in Section~\ref{sec:zsp}.}
    \label{fig:zsp-per-disease}
\end{figure}

For each patient in a disease cohort, we simulated the perturbation of relevant drug targets and measured geometrically how the resulting transcriptomic state shifted relative to healthy reference tissue. Each patient received a score between 0 and 1, where higher values indicate greater alignment toward the healthy phenotype, suggesting potential therapeutic benefit. We then computed the median score across all patients for each drug-disease combination. Drug-disease pairs were ranked by their median alignment scores, and we computed AUROC to assess discrimination between efficacious and non-efficacious treatments. We report both a global AUROC aggregated across all diseases (Table~\ref{tab:benchmark-unimodal}) and per-disease AUROC values (Figure~\ref{fig:zsp-per-disease}).

The model captured disease-specific drug effects beyond simple correlations. For example, TNF$\alpha$ inhibitors were correctly predicted as efficacious in Crohn's disease and psoriatic arthritis, but not in atopic dermatitis, reflecting the distinct physiopathology of these conditions. This context-dependent prediction contrasts with our linear baseline, built from a RNA-seq gene expression correlation matrix, which lacks such discriminative capacity (Table~\ref{tab:benchmark-unimodal}). In particular, when setting a decision threshold to 0.5, our model reaches a Positive Predictive Value of 58\%, which means that 58\% of positively predicted drugs actually worked in the disease, to be put in comparison with the 30\% success rate of phase II trials in I\&I.

\paragraph{Multimodality improves performance over separate encoders.} To evaluate the impact of multimodal post-training on downstream applications, we evaluated our embeddings on 2 downstream predictive tasks from the IBDome dataset: tissue inflammation (binary classification) and Montreal disease course classification for Crohn's disease (classification between stricturing, penetrating, or non-stricturing and non-penetrating courses). We performed the classification using the CLAM aggregation algorithm \cite{lu2021data}, either on top of EVA last layer embeddings or raw EVA-RNA and EVA-H embeddings, and show improved performance of our multimodal model (results reported in Table~\ref{tab:multimodal-evaluation}). These results suggest that post-training EVA with contrastive learning using multimodal data helped the model produce richer data representations that could be leveraged for downstream tasks.

\begin{table}[htbp]
\centering
\caption{Multimodal downstream tasks evaluation.}
\label{tab:multimodal-evaluation}
\begin{tabular}{lccc}
\toprule
\tableheaderstyle
Model & EVA & EVA-RNA + EVA-H \\
\midrule
Tissue Inflammation (AUROC) & \textbf{0.837} & 0.799\\
Montreal disease course (AUROC) & \textbf{0.585} & 0.578 \\

\bottomrule
\end{tabular}
\end{table}

\subsection{EVA-RNA integrates I\&I samples across technologies, data modalities, and species}

Transcriptomic datasets are inherently fragmented: they originate from diverse sequencing technologies (microarray, bulk RNA-seq, and single-cell RNA-seq), each with its own biases, dynamic ranges, and noise \cite{johnson2007adjusting, ritchie2015limma}. Integrating data across these technologies is highly desirable. It enables the reuse of large datasets from older technologies (e.g., microarray) while integrating both bulk RNA-seq and single-cell RNA-seq (which we treat as pseudobulk). In practice, however, such integration remains challenging \cite{tang2021rank}. Compounding this technological heterogeneity, drug development increasingly requires integration across species, especially for translational research applications. While recent foundation models such as scGPT \cite{cui2024scgpt}, Geneformer \cite{theodoris2023transfer}, BulkRNABert \cite{gelard2025BulkRNABert}, have demonstrated the capacity to learn gene, cell, or sample representations from large-scale data, they were trained exclusively on human samples and/or on one transcriptomics modality, which limits their applicability in translational research.

In this section, we investigate how EVA-RNA integrates species and technologies at multiple levels, focusing on input embeddings, contextualized gene embeddings and sample embeddings (CLS token). We demonstrate that, through joint training on microarray, bulk RNA-seq, and pseudobulked single-cell data from both human and mouse, EVA-RNA effectively learns rich representations across both species and technologies.

\begin{figure}[ht!]
    \centering
    \includegraphics[width=\textwidth]{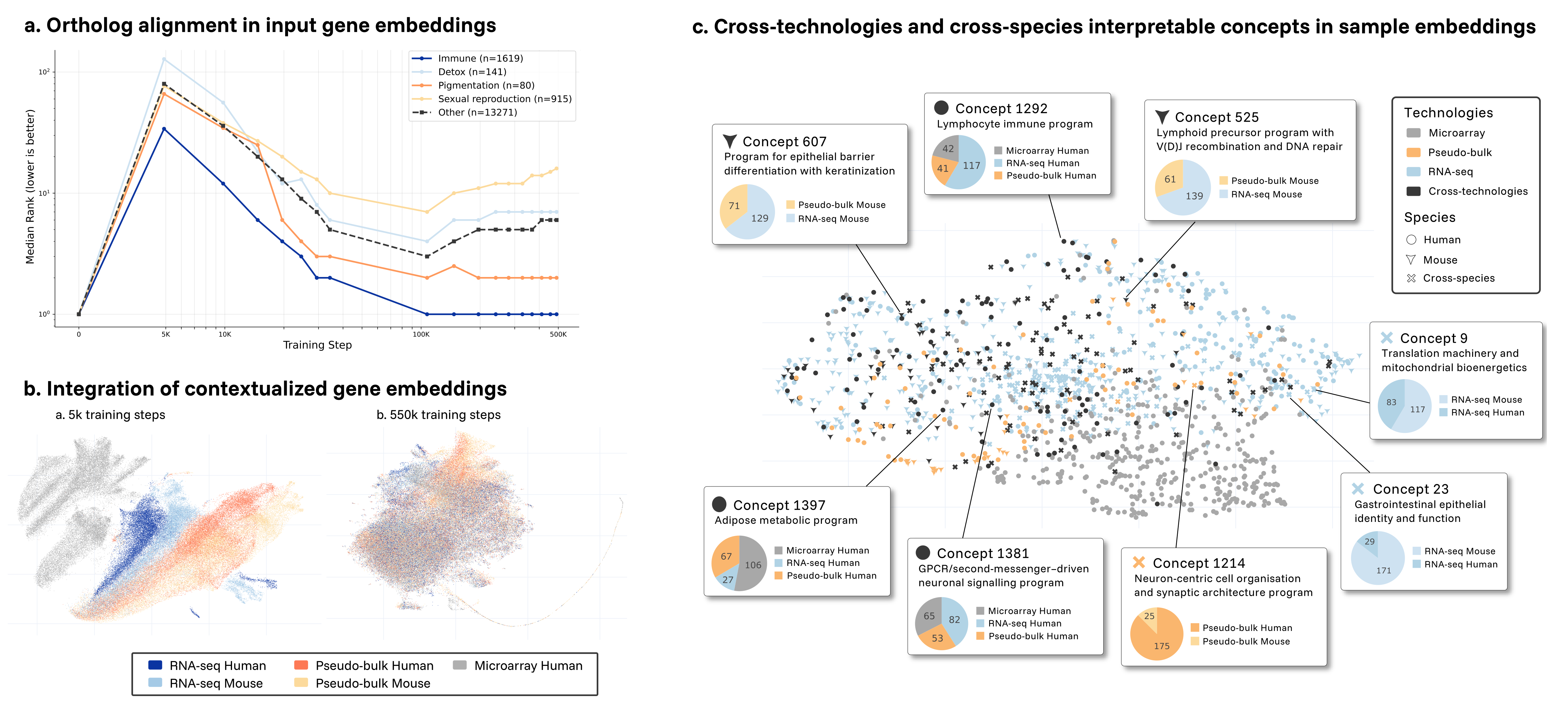}
    \caption{Cross-technologies and cross-species alignment at multiple levels in EVA-RNA. (a) Evolution of nearest neighbor median rank between orthologs based on their input embeddings. Immune genes are faster and better aligned than other groups. See Section~\ref{apx:nn-rank-evolution} for methodology details. (b) UMAP of contextualized gene embeddings from layer 30 (N-1) at 5000 training steps and 550,000 training steps. The method is described in Section~\ref{apx:context-gene-emb}. (c) UMAP of concept vectors extracted from the last CLS token of EVA-RNA with TopK sparse auto-encoder. Each point is a concept; the colors and markers correspond to the technologies and species among the 200 samples with the highest concept activation (prototypes). In boxes, we provide examples of 9 concepts, their interpretations, and the distribution of technologies and species across the 200 samples with the highest concept activations. The method is described in Section~\ref{method:sae}.}
    \label{fig:panel_alignment}
\end{figure}

\paragraph{Species alignment.} 
Figure~\ref{fig:panel_alignment}a shows the evolution of the nearest neighbor median rank in the input embedding space. Throughout training, EVA-RNA progressively aligns mouse genes with their human orthologs. We quantified this alignment using the nearest neighbor rank in all 16,168 ortholog pairs in the vocabulary. Section~\ref{apx:nn-rank-evolution} contains more details on the method. Starting from an initialization that places orthologs close together in the embedding space (Section~\ref{apx:prior-compute} explains why), the ranks initially worsen (increase) until approximately step 5,000, before steadily improving. This transient degradation could reflect the model restructuring its latent space during early training. A per-category analysis reveals that alignment quality varies across gene categories: immune genes achieve significantly lower final ranks than other groups\footnote{Bonferroni-corrected Mann-Whitney U tests showed significant differences (p < 0.05) between the Immune group and all other groups, with the exception of the Pigmentation group.}, suggesting that immunity-related genes exhibit particularly strong cross-species alignment. 

Figure~\ref{fig:panel_alignment}b shows contextualized gene embeddings from layer 30 (N-1) from an early and a late training checkpoints. Method is described in Section~\ref{apx:context-gene-emb}. Interestingly, early checkpoint clusters genes per species, while the last checkpoint shows integration of the species in a shared space.

We further applied methods from mechanistic interpretability to identify interpretable directions in the latent space of the model, which we refer to as concepts. The methods are detailed in Section~\ref{method:sae}. We trained a Top-K Sparse Auto-Encoder (topK-SAE) \cite{gaoscaling} to extract 1500 concepts from sample embeddings (using the last CLS token), and identified several concepts that detect a specific biological signal regardless of the technology or the species. 

Among the 1383 out of 1500 concepts that are active for at least 200 samples, we identified:

\begin{itemize}
    \item Single-technology and single-species concepts: 416 for human microarray, 297 for mouse RNA-seq, 200 for human RNA-seq, 75 for human pseudobulk, and 25 for mouse pseudobulk;
    \item Single-technology but cross-species concepts: 136 RNA-seq concepts for human and mouse, 3 pseudobulk concepts for human and mouse;
    \item Single-species but cross-technologies concepts: 40 mouse concepts for RNA-seq and pseudobulk, 11 human concepts for RNA-seq, pseudobulk and microarray, and 98 human concepts for two of the modalities;
    \item Cross-species and cross-technologies concepts: 82 concepts with different combinations;
\end{itemize}

Figure~\ref{fig:panel_alignment}c illustrates several concepts that we could interpret biologically. For example, concept 23 overlaps both human and mouse RNA-seq samples and can be interpreted as "gastrointestinal epithelial identity and function". Interestingly, this concept is sensitive to KRT19, GUCA2B, S100A16, PHGR1, ADH1C and their orthologs Krt19, Guca2b, S100a16, Phgr1, and Adh1, which further indicates a shared meaning in both species. Concept 1214 detects "Neuron-centric cell organization and synaptic architecture program" in both human and mouse pseudobulk samples, focusing on genes whose expression is enriched in neurons, with key players in synapse formation. 

\paragraph{Technologies alignment.} 

As described in the previous paragraph, we also observed integration of contextualized gene embeddings across technologies throughout training (Figure~\ref{fig:panel_alignment}b), as well as concepts that detect biological signals regardless of technologies (Figure~\ref{fig:panel_alignment}c). For example, concept 1292 detects "lymphocyte immune program" in human samples from all three technologies, focusing on the TCR signaling pathway (GO:0050852; TRAC, CD3D, CD3G, TRBC1, ZAP70, LCK), somatic DNA recombination in lymphocyte receptor development (GO:0016444; RAG1, RAG2, DNTT), and antigen-presentation through MHC class 1b (GO:0002475; CD1B, CD1C, CD1D, CD1E). Concept 607 detects a core tissue development program (GO:0009888) in mouse samples from all three technologies, which we term "Program for epithelial barrier differentiation with keratinization", focusing on epithelial tissue differentiation (Ovol1, Pax9, Pitx1/2), barrier remodeling/sensing (Slpi, Klk10, Klk14, Serpinb3a, Tmprss11d), and keratin production (Krt32, Krt35, Krt4, Krt76). These results suggest that the model is able to integrate different technologies in a shared space, and that it encodes meaningful and coherent biological signals.

\subsection{EVA-RNA exhibits clear pretraining scaling laws}

Whether biological foundation models for gene expression exhibit predictable scaling behavior analogous to that observed in large language models remains an open question to this day. To investigate whether scaling laws can emerge in this domain under appropriate training conditions, we conducted systematic experiments with EVA-RNA across five model sizes: 7M, 15M, 25M, 60M, and 300M parameters. All models were trained on identical data with consistent hyperparameters, varying only in the number of layers and hidden dimensions, with batch size and learning rate adapted for training stability (Table~\ref{tab:scaling-architectures}).

\begin{figure}[ht!]
    \centering
    \includegraphics[width=\textwidth]{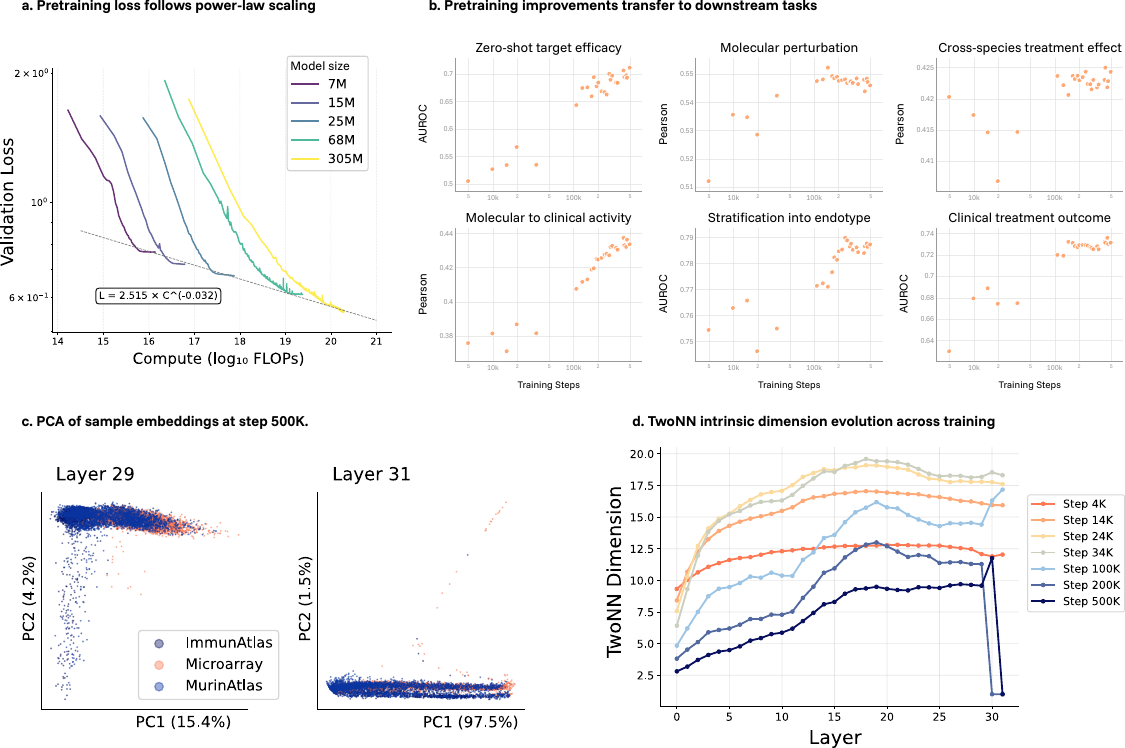}
    \caption{\textbf{EVA-RNA exhibits predictable scaling behavior across pretraining and downstream evaluation.} (a) Validation loss as a function of compute for five model sizes (7M-300M parameters). Loss follows a power-law relationship with no evidence of plateau. (b) Downstream task performance as a function of training steps across six evaluation categories, showing a clear improvement with continued pretraining. (c) PCA of sample embeddings at layers 29 (top) and 31 (bottom), colored by data source. Layer 29 (N-2) retains multi-dimensional structure, while layer 31 (N) collapses onto the first principal component, reflecting compression toward the sample-level reconstruction objective. (d) TwoNN intrinsic dimension across transformer layers at different training checkpoints. Early layers maintain high-dimensional representations throughout training, while later layers progressively compress the contextualized gene representations. This compression effect intensifies with training, with final layer showing increasingly sharp rank reduction at later checkpoints. See section~\ref{apx:intrisic-dimensionality} for more details.}
    \label{fig:scaling-laws}
\end{figure}

Our scaling experiments reveal that EVA-RNA follows predictable power-law scaling behavior across model sizes (Figure~\ref{fig:scaling-laws}a). Fitting a power law of the form $L = aC^{-b}$ to the validation loss as a function of compute yields $L = 2.515 \times C^{-0.032}$, indicating that each order of magnitude increase in compute reduces validation loss by approximately 7\%. Critically, we observe no evidence of a plateau at the 300M parameter scale, suggesting that further scaling may yield continued improvements. This finding contrasts with prior work on single-cell foundation models: AIDO.Cell reported diminishing returns beyond 100M parameters~\cite{ho2024scaling}, raising questions about whether gene expression models could benefit from scale. Our results suggest that appropriate training data curation (cross-species, multi-technology) and architectural choices may be necessary conditions for observing scaling laws in this domain.

Analysis of intermediate representations reveals a characteristic compression pattern across transformer layers (Figure~\ref{fig:scaling-laws}c--d). PCA of sample embeddings shows that layer 29 (N-2) retains multi-dimensional structure with variance distributed across multiple principal components, while layer 31 collapses onto the first principal component, concentrating 97.5\% of variance. This compression intensifies with training: TwoNN intrinsic dimensionality (ID) \cite{facco2017estimating} across layers reveals that early and middle layers gradually reorganize their representations as training proceeds in two successive dynamics (Figure~\ref{fig:scaling-laws}d):

\begin{itemize}
    \item Expansion phase: during the early steps (sub-100k), gene token representations are progressively enriched and a first gene space is learnt; we see the representations ID steadily increasing across the network during this phase.
    \item Compression phase: later in the training (after step 100k) and under the pressure of regularization, the model progressively learns more efficient gene tokens representations, as highlighted by the steady decrease of the ID over the later steps -- while training and validation losses keep decreasing.
\end{itemize}

The final transformer layer undergoes a sharp collapse especially in later checkpoints, dropping to an intrinsic dimension around 1 at step 500K -- a compression that becomes more pronounced throughout training. This pattern reflects the objectives in EVA-RNA pretraining: earlier layers learn increasingly rich, distributed gene-gene relationships, while the final layer specializes for the gene expression reconstruction objective, compressing representations onto a low-dimensional manifold suitable for one-shot expression decoding.

Importantly, pretraining improvements transfer to downstream task performance across all evaluation categories (Figure~\ref{fig:scaling-laws}d), though scaling profiles vary across tasks. Zero-shot target efficacy, clinical activity prediction, and clinical treatment outcome show sustained improvement throughout training. Molecular perturbation and cross-species treatment effect show early gains but more modest continued improvement, with higher variance. This pattern suggests that tasks requiring patient-level representations (treatment response, clinical severity) benefit most from extended pretraining, while perturbation prediction tasks may extract most of their value from earlier training.

\begin{table}[htbp]
\centering
\caption{EVA-RNA model architectures used in scaling experiments, with their corresponding pretraining data regimes.}
\label{tab:scaling-architectures}
\begin{tabular}{@{}lrrrrrrrr@{}}
\toprule
\tableheaderstyle Model & Layers & Heads & Hidden & Head size & FFN & Vocab embed & Batch size & LR \\
\midrule
7M   & 1  & 2  & 64  & 32 & 128  & 64  & 200k tokens & $2.5 \times 10^{-3}$ \\
15M  & 4 & 2  & 128 & 64 & 512 & 64 & 240k tokens & $3.0 \times 10^{-4}$ \\
25M  & 4 & 8  & 256 & 32 & 512 & 64 & 240k tokens & $3.0 \times 10^{-4}$ \\
60M  & 24 & 8  & 256 & 32 & 2048 & 128 & 240k tokens & $3.0 \times 10^{-4}$ \\
300M & 32 & 12 & 768 & 64 & 3072 & 256 & 660k tokens & $1.2 \times 10^{-4}$ \\
\bottomrule
\end{tabular}
\end{table}

\section{Methods}

\subsection{Datasets}

\subsubsection{RNA expression datasets} \label{sec:rna-datasets}

EVA-RNA was pretrained on ImmunAtlas, a gene expression atlas sourced from public I\&I datasets containing a total of 545,343 samples spanning mouse and human samples, multiple technologies, and platforms. The corpus comprises five complementary datasets curated for immunology research (Table~\ref{tab:datasets}). All datasets underwent QA/QC pipelines and log-normalization: counts were normalized to counts per million (not applied to microarray samples), then were log-transformed via $\log_2(x + 1)$. During training, datasets are sampled with weights emphasizing bulk RNA-seq while incorporating cross-species and single-cell-derived signals, for a total of 330B gene tokens; Table~\ref{tab:datasets} references the effective number of epochs per dataset.

\begin{table}[htbp]
\centering
\caption{pretraining dataset composition. Weight indicates the relative sampling probability during training; higher weights increase the frequency at which samples from that dataset are drawn, emphasizing bulk RNA-seq data while maintaining cross-species and single-cell representation. *One gene token = one gene name and corresponding expression value. **Effective epochs computed for 330B total training tokens.}
\label{tab:datasets}
\begin{tabular}{lccccccc}
\toprule
\tableheaderstyle
Dataset & Species & Technology & Samples & Genes & Tokens* & Weight & Eff. Epochs** \\
\midrule
ImmunAtlas-seq & Human & Bulk RNA-seq & 42,166 & 39,376 & 1.7B & 0.30 & 58.2 \\
ImmunAtlas-MA & Human & Microarray & 55,564 & 39,376 & 2.2B & 0.20 & 30.0 \\
MurinAtlas & Mouse & Bulk RNA-seq & 437,899 & 26,864 & 11.8B & 0.40 & 11.2 \\
CellxGene human & Human & Pseudobulk & 8,498 & 39,376 & 334M & 0.05 & 49.4 \\
CellxGene mouse & Mouse & Pseudobulk & 1,216 & 26,864 & 33M & 0.05 & 500.0 \\
\midrule
\textbf{Total} & - & - & 545,343 & 66,240 & 16.1B & 1.00 & 20.5 \\
\bottomrule
\end{tabular}
\end{table}

\paragraph{Gene Vocabulary.} EVA-RNA uses a multi-species gene vocabulary consisting of 66,240 human and mouse NCBI Gene IDs -- chosen over gene symbols to avoid ambiguity from synonyms and naming inconsistencies across data sources. The vocabulary was filtered to only contain genes present in the bulk RNA-seq datasets, excluding the genes that appear only in single-cell or microarray data. This filtering ensures that all genes in the vocabulary have sufficiently high-quality training examples from quantitatively reliable bulk RNA-seq measurements, avoiding genes with sparse or potentially biased expression estimates from other platforms. Each gene is assigned a unique token index, with additional special tokens reserved for CLS, MASK, and PAD operations.

\subsubsection{Histology datasets}

The histology training data comprises 4,076 whole-slide images (WSIs) yielding approximately 20 million tissue tiles (224 $\times$ 224 images also called "patches") from 1,252 patients across five curated datasets. Slides are stained with hematoxylin and eosin (H\&E), with a subset including CD3 immunohistochemistry (IHC).

\begin{table}[ht]
\centering
\caption{Overview of histology training data.}
\label{tab:histo_data}

\begin{subtable}[c]{0.48\linewidth}
\centering
\caption{Disease coverage}
\resizebox{\linewidth}{!}{
\begin{tabular}{lccc}
\toprule
\tableheaderstyle Condition & Patients & Slides & Tiles (M) \\
\midrule
Crohn's disease      & 367 & 843 & 6.8 \\
Ulcerative colitis   & 216 & 611 & 5.8 \\
Sjögren's disease    & 285 & 1,488 & 4.5 \\
Control     & 224 & 858 & 3.1 \\
Other                & 160 & 276 & 0.2 \\
\midrule
\textbf{Total}       & \textbf{1,252} & \textbf{4,076} & \textbf{20.4} \\
\bottomrule
\end{tabular}
}
\end{subtable}
\hfill
\begin{subtable}[c]{0.48\linewidth}
\centering
\caption{Tissue coverage}
\resizebox{\linewidth}{!}{
\begin{tabular}{lccc}
\toprule
\tableheaderstyle Tissue & Patients & Slides & Tiles (M) \\
\midrule
Colon                        & 657 & 1,119 & 8.5 \\
Salivary gland               & 435 & 2,232 & 6.8 \\
Small intestine (ileum)      & 502 & 613 & 3.9 \\
Esophagus                    & 17 & 33 & 0.3 \\
Stomach                      & 16 & 17 & 0.3 \\
Rectum                       & 21 & 21 & 0.1 \\
Other                        & 39 & 41 & 0.6 \\
\bottomrule
\end{tabular}
}
\end{subtable}

\end{table}
\paragraph{Disease \& tissue coverage.} Samples include inflammatory bowel disease (IBD) cases, such as Crohn's disease, Ulcerative colitis, alongside Sjögren's Disease and healthy controls. This provides balanced representations across autoimmune and inflammatory conditions. 

The corpus spans 10 tissue categories, such as colon, salivary gland, small intestine (including ileum), stomach, esophagus, and rectum. These cover the major anatomical sites relevant to I\&I diseases. See Table~\ref{tab:histo_data} for more details.

\begin{figure}[ht!]
     \centering
     \begin{subfigure}[b]{0.48\textwidth}
         \centering
         \includegraphics[width=\textwidth]{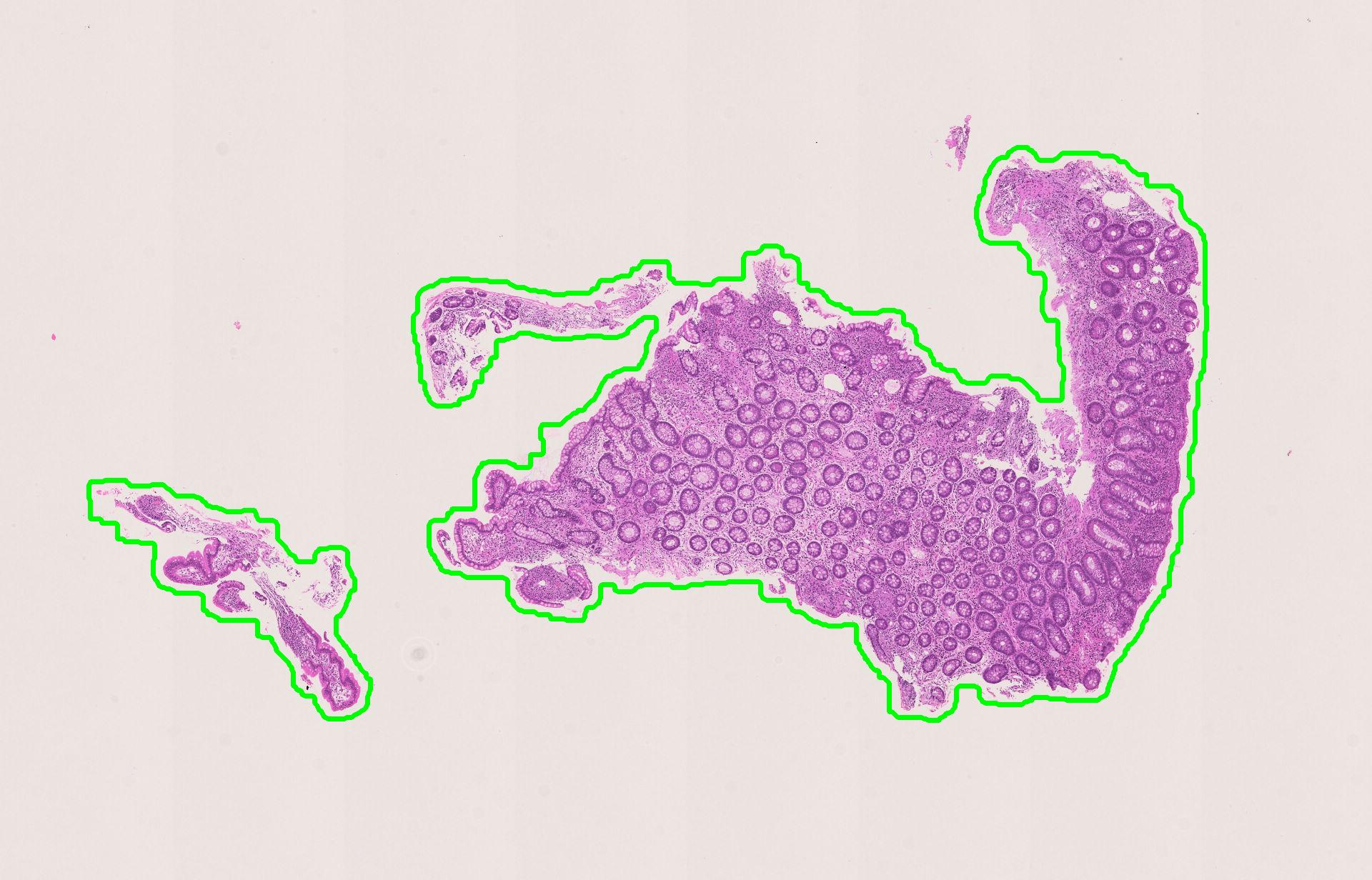}
         \caption{Tissue segmentation}
         \label{fig:segmentation}
     \end{subfigure}
     \hfill
     \begin{subfigure}[b]{0.48\textwidth}
         \centering
         \includegraphics[width=\textwidth]{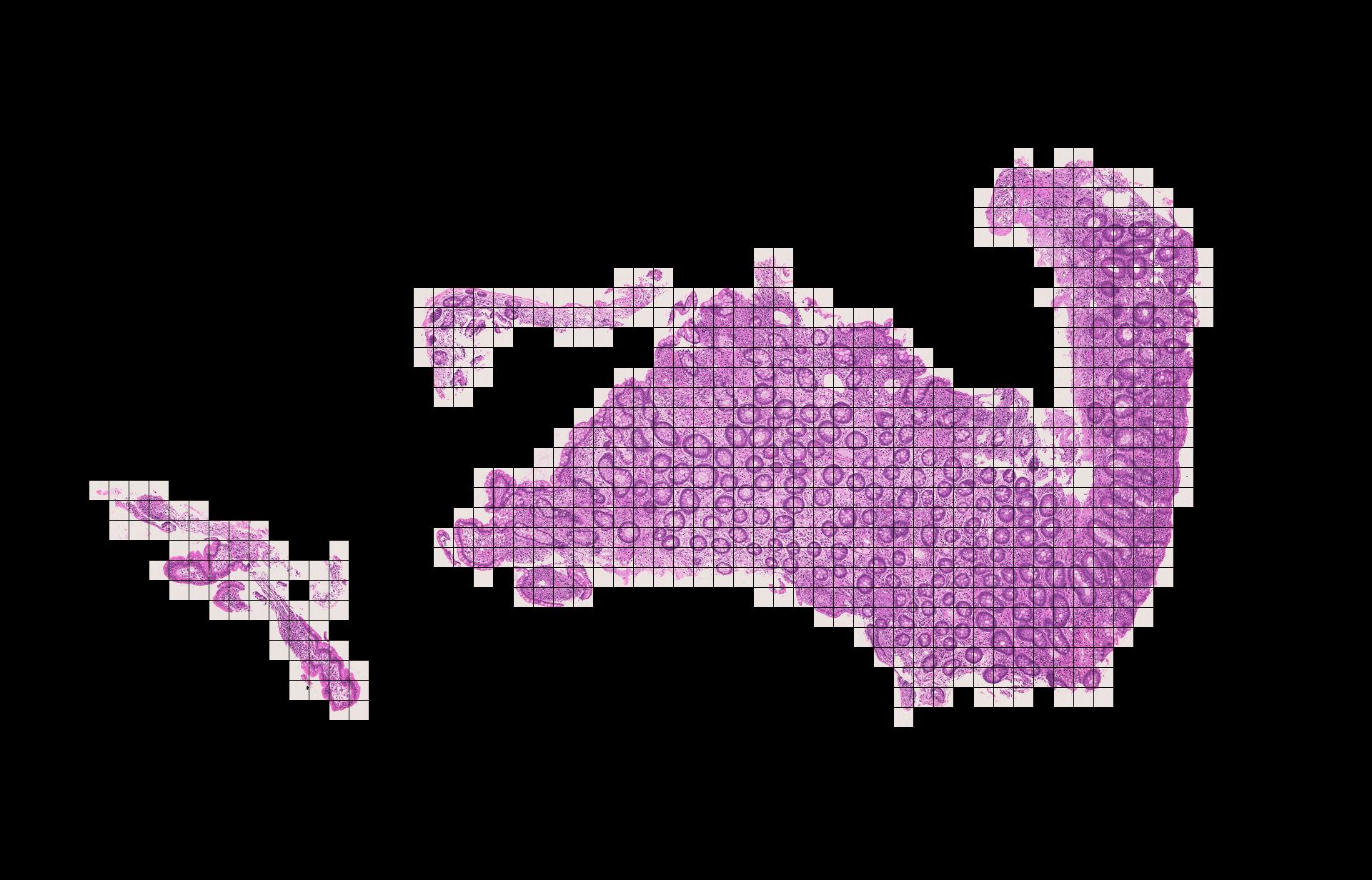}
         \caption{Tiles extraction}
         \label{fig:tiling}
     \end{subfigure}
     \caption{WSI Preprocessing: From initial tissue segmentation to the extraction of localized tiles for model input.}
     \label{fig:wsi-preprocessing}
\end{figure}

\paragraph{Preprocessing}
As shown in Figure~\ref{fig:wsi-preprocessing}, WSIs are first segmented (to keep tissue only) then cut into small tiles of size 224 $\times$ 224. Each tile is then passed through the model as an input. Preprocessing details can be found in Appendix~\ref{apx:wsi-preprocessing}.

\subsection{EVA-RNA encoder} \label{sec:eva-rna-encoder}

\subsubsection{EVA-RNA encoder architecture}

The EVA-RNA encoder follows a transformer architecture with 32 layers (indexed from 0 to 31 throughout the paper), 768 hidden dimensions, 12 attention heads, and 3072-dimensional feedforward layers, totaling 305 million parameters. We employ pre-layer normalization with residual scaling by $1/\sqrt{2L}$ where $L$ is the number of layers, following established practices for training deep transformers \cite{radford2019language}.

Gene expression values are embedded into the hidden space through a multi-layer perceptron with architecture $[1 \rightarrow 16 \rightarrow 128 \rightarrow 384 \rightarrow 768]$ followed by layer normalization. The final gene representation is the sum of the gene identity embedding and value embedding.

Each input sequence is prepended with a CLS token whose final representation serves as the sample-level embedding. MASK tokens replace masked gene positions during pretraining, and PAD tokens handle variable-length sequences.

\subsubsection{Gene embeddings initialization with external knowledge}

Inspired by other transcriptomics foundation model approaches \cite{yang2024genecompass,kalfon2025scprint}, EVA-RNA leverages external knowledge via precomputed gene embeddings coming from five different sources.

\begin{itemize}
    \item scGPT gene embeddings capturing single-cell co-expression patterns \cite{cui2024scgpt}
    \item ESM-2 (650M) protein embeddings encoding amino acid sequence information \cite{lin2023evolutionary}
    \item Text embeddings extracted from NCBI gene descriptions \cite{sayers2025database}
    \item Text embeddings extracted from UniProt protein description \cite{bateman2025uniprot}
    \item Knowledge graph derived embeddings (KGE) leveraging RotatE method \cite{sun2019rotateknowledgegraphembedding}
\end{itemize}

\noindent By integrating external knowledge from diverse sources, including text, knowledge graphs, and foundation models that capture transcriptomic, proteomic, and functional biological information, our goal is to initialize EVA-RNA gene embeddings with rich, biologically informed representations that can then be further refined during training. See Appendix~\ref{apx:prior-compute} for further details about external knowledge embeddings computation.

\begin{figure}[!ht]
    \centering
    \includegraphics[width=\textwidth]{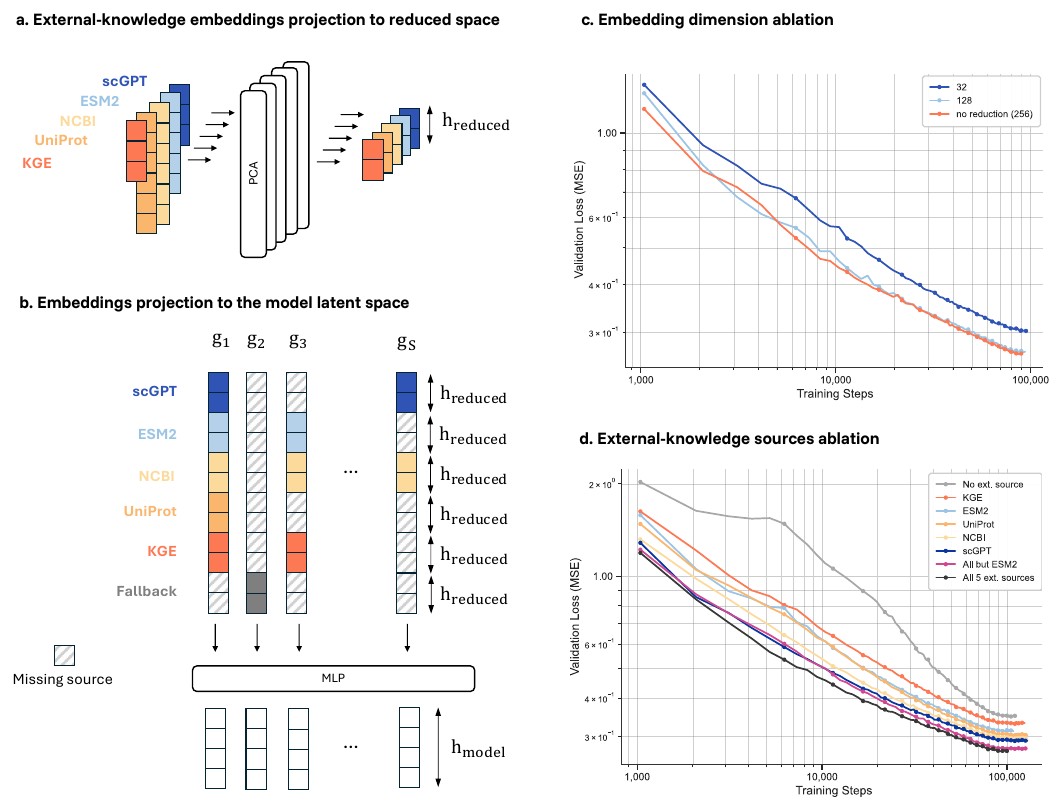}
    \caption{(a) At initialization, all external knowledge embeddings are reduced to a shared dimension via a per-source PCA. (b) For a given gene $g_{i}$, reduced embeddings from all sources are concatenated (channels are zeroed out if $g_{i}$ is not supported by the source, e.g., $g_{3}$ is not supported by scGPT and UniProt) and passed through an MLP to generate the final embedding. On the figure, $g_{2}$ is not supported by any external knowledge source, hence its embedding is only derived from the fallback matrix. (c) Using a reduced vocabulary size (128) provides similar training dynamics as full-size (256) while saving training parameters. See Appendix~\ref{apx:emb-dimension-ablation} for more details. (d) Using all 5 external knowledge sources yields the best performance both in convergence and final validation loss. See Appendix~\ref{apx:prior-ablation} for more details.}
    \label{fig:prior-panel}
\end{figure}

Figure~\ref{fig:prior-panel}a-b details how these external knowledge embeddings are merged and provided to the model.
First, the $N_{sources}$ external knowledge matrices are reduced at initialization using a per-source PCA. This ensures embeddings coming from the different external sources are all of the same size $h_{reduced}$. A global fallback matrix (initialized at random) is added to account for genes that are not supported by any external sources.  After initialization, these $N_{sources} + 1$ embedding matrices are all trainable parameters, updated jointly with the rest of the model during training.
To retrieve the embedding of a gene $g_{i}$, the $N_{sources} + 1$ embedding matrices are queried. If the gene is contained in the matrix, its embedding is fetched. Otherwise, a null vector is used. Then, the $N_{sources} + 1$ embeddings are concatenated into a single embedding of size $(N_{sources} + 1)\cdot h_{reduced}$ and passed through a two-layer MLP (with an expansion factor of 4) to retrieve an embedding of size $h_{model}$ that is ready to be forwarded to the encoder. We use $h_{reduced}$ < $h_{model}$ to save parameters, making the model significantly smaller hence faster to train. 

We conducted an ablation study on a smaller model ($\sim$55M parameters) showing there was no real performance difference between keeping the same dimension and reducing the embedding size by a factor 2 (see Figure~\ref{fig:prior-panel}c). For the 300M model of hidden size 768, we use a reduced size of 256, which results in a saving of 108M trainable parameters i.e., $\sim$37\% of the current model size.

We also conducted an ablation study (see Figure~\ref{fig:prior-panel}d) on a smaller model ($\sim$55M parameters) to ensure that combining all 5 external knowledge sources helped the model learn better and faster. 

\subsubsection{Distributional gene expression decoder}

Single-cell RNA-seq data exhibit characteristic statistical properties, including a high degree of sparsity due to technical dropout and biological zeros, as well as overdispersion relative to the Poisson distribution. To capture these properties, EVA employs a Zero-Inflated Negative Binomial (ZINB) decoder that models the full conditional distribution of gene expression, which is a standard gene expression modeling paradigm notably used in scVI \cite{lopez2018deep} and scPRINT \cite{kalfon2025scprint}. For bulk and pseudobulk RNA-seq, where zero-inflation is less pronounced, the model can effectively reduce to a standard Negative Binomial.

The ZINB distribution models expression $x \geq 0$ as a mixture of a point mass at zero and a Negative Binomial,

\begin{equation}
P(X = x) = \begin{cases}
\pi + (1-\pi) \cdot \text{NB}(0; \mu, \theta) & x = 0 \\
(1-\pi) \cdot \text{NB}(x; \mu, \theta) & x > 0
\end{cases}
\end{equation}

where the Negative Binomial PMF is defined with respect to mean ($\mu$) and dispersion ($\theta$) parameters,

\begin{equation}
\text{NB}(x; \mu, \theta) = \frac{\Gamma(x + \theta)}{\Gamma(\theta) \cdot \Gamma(x+1)} \left(\frac{\theta}{\theta + \mu}\right)^\theta \left(\frac{\mu}{\theta + \mu}\right)^x
\end{equation}

For each masked gene expression position, the decoder predicts three parameters from the contextualized token embedding $\mathbf{h} \in \mathbb{R}^d$ via linear projections,

\begin{align}
\mu &= \text{softplus}(\mathbf{w}_\mu^\top \mathbf{h}) & \text{(mean)} \\
\theta &= \text{softplus}(\mathbf{w}_\theta^\top \mathbf{h}) & \text{(dispersion)} \\
\pi &= \sigma(\mathbf{w}_\pi^\top \mathbf{h}) & \text{(zero-inflation)}
\end{align}

where $\text{softplus}(x) = \log(1 + e^x)$ ensures positivity, and $\sigma$ is the sigmoid function.

% \paragraph{Weights initialization.} All EVA-RNA's attention matrices and feed-forward networks are initialized with the default Kaiming uniform initialization. Moreover, weights of residual layers are scaled at initialization by a factor of $1/{\sqrt{2L}}$ with $L$ being the number of layers in our transformer, following established practices for training deep transformers.

% DON'T DELETE
%Default Kaiming Uniform initialization (default initialization for nn.Linear) is used for all our FFN blocks. For Attention blocks, it depends on whether we use flash or not. If flash attention is used, Kaiming uniform is used (cf \href{https://github.com/Dao-AILab/flash-attention/blob/f472175a9e9ecbd7178d24e5f3eb3cc7925b1173/flash_attn/modules/mha.py#L459}{here} and \href{https://github.com/Dao-AILab/flash-attention/blob/f472175a9e9ecbd7178d24e5f3eb3cc7925b1173/flash_attn/modules/mha.py#L481}{here}). But if flash is not used, Xavier normal is used (cf \href{https://github.com/pytorch/pytorch/blob/045e8f98cdd31fe0747a5e798d75e1d6c7602a74/torch/nn/modules/activation.py#L1236}{here}).

\subsubsection{Gene-level training objective.}

The gene-level pretraining objective minimizes the negative log-likelihood over masked positions,

\begin{equation}
\mathcal{L}_{\text{ZINB}} = -\frac{1}{|\mathcal{M}|} \sum_{i \in \mathcal{M}} \log P(x_i \mid \mu_i, \theta_i, \pi_i)
\end{equation}

where $\mathcal{M}$ denotes the set of masked gene indices. The log-probability is computed using the log-gamma function for numerical stability.

The ZINB formulation provides several advantages over standard mean squared error (MSE) reconstruction losses. The zero-inflation parameter $\pi$ explicitly models the dual origins of zeros in single-cell RNA-seq data, i.e., biological zeros (genes genuinely not expressed) and technical dropouts, whereas the MSE loss treats all zeros equivalently. The dispersion parameter $\theta$ captures the  overdispersion characteristic of count data, where variance exceeds the mean following $\text{Var}(X) = \mu + \mu^2/\theta$; in contrast, MSE implicitly assumes homoscedastic Gaussian noise, leading to systematic underweighting of lowly-expressed genes. Finally, the distributional output enables principled uncertainty quantification in downstream applications, as the predicted parameters define a full probability distribution from which confidence intervals can be derived, rather than merely providing point estimates.

\subsubsection{CLS token reconstruction}

The CLS token embedding $\mathbf{h}_{\text{CLS}} \in \mathbb{R}^d$ is trained through an auxiliary objective that enforces global sample-state awareness. Given a batch of $N$ samples, each containing $G$ genes with expression values $\mathbf{x} = (x_1, \ldots, x_G)$, the model learns to reconstruct all gene expressions using only the contextualized CLS embedding and the input (i.e. non-contextualized) gene embeddings $\mathbf{E}_g \in \mathbb{R}^{G \times d}$. Specifically, the predicted expression profile is computed as:
  \begin{equation}
  \hat{\mathbf{x}} = f_{\text{decode}}(\mathbf{h}_{\text{CLS}}, \mathbf{E}_g)
  \end{equation}
  where $f_{\text{decode}}$ is a decoder network (a two-layer MLP with an increase in width of factor 4) that combines the sample-level representation from the CLS token with gene-specific information from the gene embeddings. The CLS loss is then defined as:
  \begin{equation}
  \mathcal{L}_{\text{CLS}} = \frac{1}{N \cdot G} \sum_{i=1}^{N} \sum_{j=1}^{G} \left( \hat{x}_{ij} - x_{ij} \right)^2
  \end{equation}
  This objective is combined with the primary ZINB task loss (masked gene expression prediction $\mathcal{L}_{\text{MGE}}$ or denoising $\mathcal{L}_{\text{denoise}}$) via a weighted sum:
  \begin{equation}
  \mathcal{L}_{\text{total}} = (1 - \lambda) \mathcal{L}_{\text{ZINB}} + \lambda \mathcal{L}_{\text{CLS}}
  \end{equation}
  where $\lambda \in [0, 1]$ controls the trade-off. The key intuition is that forcing the CLS token to reconstruct the entire expression profile encourages it to learn a compressed, holistic representation of the sample state that captures the global transcriptional program. Unlike the primary task, which focuses on local gene-to-gene relationships through masked prediction, the CLS objective ensures the model maintains awareness of the overall transcriptional phenotype, allowing the model to learn both specific and global components simultaneously. This dual objective prevents the model from overfitting to local patterns while improving downstream tasks that require transcriptomic profile-level information such as clinical activity or treatment outcome predictions.

\subsubsection{Data Augmentation}

EVA-RNA is pretrained using a masked gene expression (MGE) task analogous to masked language modeling. For each input profile, a set fraction of gene expression values are replaced with a MASK token. The model has to predict these masked expression values using the other non-masked gene tokens. We employ a curriculum learning schedule, a methodology explored in NLP \cite{wettig2023should,ankner2024dynamic,yoon2025currimae}, where the masking ratio decreases linearly from 95\% to 15\% over 500,000 training steps, enabling the model to learn first from highly contextualized predictions before transitioning to scenarios with richer input context (see Appendix~\ref{apx:masking-ratio}).

To improve training efficiency and enable the model to handle variable-length inputs, we implemented a block size expansion schedule. The maximum sequence length increases linearly from 600 to 1,800 genes over 400,000 steps following a 1,000-step warm-up period. Validation is performed with a fixed block size of 1,200 genes.

We apply mix-up augmentation to all samples with $\alpha = 1.0$, linearly interpolating expression profiles between randomly paired samples within each batch. We found that this aggressive mix-up setting helped the model generalize across technologies and species.

\subsubsection{Training run}

We use AdamW optimization with $\beta_1 = 0.9$, $\beta_2 = 0.999$, and weight decay 0.05. The learning rate follows a warm-up-cosine schedule: linear warm-up to $1.7 \times 10^{-4}$ over 1,000 steps, then cosine decay to $5 \times 10^{-6}$ over 250,000 steps. Gradient norms are clipped to 1.0. Training uses mixed-precision (bfloat16) with gradient accumulation over 2 steps and batch size 32, yielding an effective batch size of 64.

The model was trained for approximately 4,000 hours on A100 GPUs using distributed data parallel (DDP) training across multiple nodes, with bfloat16 mixed precision and TF32 matmul precision.

\subsection{EVA-H encoder} \label{sec:eva-h}

Our histology encoder is an 86M-parameter ViT-B/14-based vision transformer. We use Hibou-B \cite{nechaev2024hibou}  to initialize its weights. Hibou-B was pretrained on $>$1M histology whole slide images. The model processes 224$\times$224 pixel images with a 14$\times$14 patch size.

\subsubsection{EVA-H training}
\paragraph{Training objective.}
Our histology foundation model is trained using the iBOT \cite{zhou2021ibot} self-supervised learning framework, which combines two complementary objectives through a teacher-student architecture. The student network processes both global and local multi-scale crops of histology tiles, while an exponential moving average teacher network provides stable training targets. The training objective consists of: (1) a DINO loss that enforces consistency between teacher and student class token predictions across different augmented views using cross-entropy with temperature scaling and centering, and (2) a masked image modeling (MIM) loss that reconstructs masked tile tokens by minimizing the cross-entropy between student predictions and detached teacher targets on randomly masked patches. The final loss is computed as a weighted combination:
\begin{equation}
\mathcal{L} = (1 - \lambda)\mathcal{L}_{\text{MIM}} + \lambda\mathcal{L}_{\text{DINO}}
\end{equation}
where $\lambda$ controls the relative contribution of each objective. The teacher parameters are updated via momentum ($m = 0.996$) without gradient propagation, ensuring stable and consistent pseudo-labels throughout training.

\paragraph{Training hyperparameters.}
The model is fine-tuned using Low-Rank Adaptation \cite{hu2021lora} with rank $r=8$, scaling parameter $\alpha=16$, and no dropout, enabling parameter-efficient training while maintaining model expressivity. Training is conducted for 7 epochs during 90 hours on 4 Nvidia H100 GPUs with a batch size of 64 tiles per device, using the AdamW optimizer \cite{loshchilov2017decoupled} with an initial learning rate of $2 \times 10^{-5}$ and weight decay of $1 \times 10^{-5}$. The learning rate follows a step-decay schedule with $\gamma=0.9$ applied every 4,000 optimization steps. For the iBOT framework, we employ an exponential moving average (EMA) teacher with momentum $\tau_t=0.996$ and center momentum $\tau_c=0.9$, using asymmetric temperature sharpening with $T_s=0.1$ for the student and $T_t=0.04$ for the teacher networks. The pretraining strategy generates 2 global crops and 2 local crops per image. Stochastic masked token prediction is applied to the first global crop with 50\% probability to balance the self-distillation and reconstruction objectives. This stochasticity, as established in the iBOT framework, stabilizes training by mitigating the distribution mismatch between masked global and unmasked local crops \cite{zhou2021ibot}. Specifically, when masking is active, tokens are randomly masked using block-wise masking with masking ratios sampled uniformly between 0.1 and 0.5, while the remaining views are used for the DINO loss. The masked image modeling objective is balanced with the DINO loss using $\lambda=0.5$. Training utilizes mixed precision (bfloat16) for computational efficiency and gradient clipping with a maximum norm of 1.0 to ensure training stability.

\subsection{Multimodal fusion model}

We develop a transformer-based fusion architecture that integrates transcriptomics data and histology whole slide images into a unified embedding space. The model consists of the frozen pretrained unimodal encoders EVA-RNA and EVA-H, followed by a learnable fusion transformer that performs cross-modal attention (Figure~\ref{fig:global-architecture}).

\subsubsection{Multimodal tokenization}

The fusion architecture treats each modality's representations as a sequence of tokens that attend to one another via standard self-attention. We project the unimodal representations into a shared $d$-dimensional joint input space using modality-specific linear projections previously learned:

\begin{align}
    \mathbf{z}_{\text{rna-cls}} &= \mathbf{W}_{\text{rna-cls}} \mathbf{h}_{\text{cls}} + \mathbf{b}_{\text{rna-cls}} \\
    \mathbf{z}_{\text{gene},i} &= \mathbf{W}_{\text{rna-gene}} \mathbf{h}_{\text{gene},i} + \mathbf{b}_{\text{rna-gene}} \quad \text{for } i = 1, \ldots, n_{\text{genes}} \\
    \mathbf{z}_{\text{tile},j} &= \mathbf{W}_{\text{histo}} \mathbf{h}_{\text{tile},j} + \mathbf{b}_{\text{histo}} \quad \text{for } j = 1, \ldots, n_{\text{tiles}}
\end{align}

\noindent where $\mathbf{h}_{\text{cls}}$ and $\mathbf{h}_{\text{gene},i}$ are the RNA encoder's CLS and gene token outputs, $\mathbf{h}_{\text{tile},j}$ are the histology tile embeddings. The complete input sequence to the fusion transformer is,

\begin{equation}
    \mathbf{X} = [\mathbf{z}_{\text{cls}}; \mathbf{z}_{\text{rna-cls}}; \mathbf{z}_{\text{gene},1}; \ldots; \mathbf{z}_{\text{gene},n}; \mathbf{z}_{\text{tile},1}; \ldots; \mathbf{z}_{\text{tile},m}]
\end{equation}

\noindent where $\mathbf{z}_{\text{cls}}$ is a learnable CLS token whose contextualized output serves as the final joint embedding.

To enable flexible inference when one modality is absent, we introduce learnable fallback embeddings $\mathbf{e}_{\text{rna}} \in \mathbb{R}^d$ and $\mathbf{e}_{\text{histo}} \in \mathbb{R}^d$ initialized from $\mathcal{N}(0, 0.02)$. So that, for instance, when RNA data for a sample are missing (resp. histology), $\mathbf{e}_{\text{rna}}$ is the only RNA token used.

\subsubsection{Model architecture}

The fusion transformer consists of 6 standard transformer encoder layers with 8 attention heads, $d = 768$  hidden dimensions, and a feed-forward dimension $4d = 3072$. We use pre-layer normalization, GELU activations, a dropout probability 0.1, and Flash Attention 2. The output at the CLS position, after a final layer normalization, constitutes the joint multimodal embedding,

\begin{equation}
    \mathbf{z}_{\text{joint}} = \text{LayerNorm}\left(\text{Transformer}(\mathbf{X})_0\right)
\end{equation}

\noindent We have not yet conducted extensive, rigorous parameter searches and ablation studies on this multimodal transformer architecture. 

\subsection{Contrastive pretraining}

We train the fusion model using contrastive learning, where views derived from the same biological sample are pulled together while views from different samples are pushed apart.

\subsubsection{Multi-View Generation}

For each sample, we generate multiple augmented views through stochastic subsampling:

\begin{enumerate}
    \item \textbf{RNA views}: Subsampled gene set (1200 gene tokens), with histology masked by the fallback embedding
    \item \textbf{Histology views}: Subsampled tiles (256 to 1024) with RNA masked by the fallback embedding
    \item \textbf{Multimodal views}: Subsampled genes combined with subsampled tiles
\end{enumerate}

This view generation strategy encourages the model to learn representations that are invariant to the specific subset of genes or tiles observed, while also enabling cross-modal alignment by treating different modality combinations from the same sample as positive pairs.

\subsubsection{Multi-Positive InfoNCE Loss}

We employ a multi-positive variant of the InfoNCE \cite{oord2018representation,khosla2020supervised} loss that accommodates multiple positive pairs per anchor. Let $\mathcal{V} = \{(\mathbf{z}_i, s_i)\}_{i=1}^{N}$ be the set of $N$ view embeddings in a batch, where $s_i$ denotes the sample index for view $i$. For anchor $i$, the positive set is $\mathcal{P}(i) = \{j: s_j = s_i, j \neq i\}$. The loss is:
\begin{equation}
    \mathcal{L} = -\frac{1}{|\mathcal{A}|} \sum_{i \in \mathcal{A}} \left[ \log \frac{1}{|\mathcal{P}(i)|} \sum_{p \in \mathcal{P}(i)} \exp\left(\frac{\text{sim}(\mathbf{z}_i, \mathbf{z}_p)}{\tau}\right) - \log \sum_{j \neq i} \exp\left(\frac{\text{sim}(\mathbf{z}_i, \mathbf{z}_j)}{\tau}\right) \right]
\end{equation}
where $\mathcal{A} = \{i: |\mathcal{P}(i)| > 0\}$ is the set of anchors with at least one positive, $\text{sim}(\mathbf{u}, \mathbf{v}) = \mathbf{u}^\top \mathbf{v} / \|\mathbf{u}\| \|\mathbf{v}\|$ is cosine similarity, and $\tau = 0.1$ is the temperature. This formulation, which averages over positives in the numerator, generalizes the standard InfoNCE to handle multiple positives per anchor.

\subsubsection{Training parameters}

Training is performed for 500 hours on A100 GPUs, using bfloat16 mixed precision and TF32 matrix multiplication, with an AdamW optimizer (weight decay $10^{-2}$, momentum parameters $\beta_1 = 0.9$, $\beta_2 = 0.99$), and a warm-up-cosine schedule for the learning rate (peak LR at $3 \times 10^{-4}$). Gradient norm clipping was used with a maximum value of 1. The batch size used is 16 samples per GPU, with 2 views generated per sample per modality type, yielding up to 96 views per batch when all modality combinations are present. Per view, maximum sequence lengths are 1,200 gene tokens for transcriptomics, and 512 to 4,096 histology tiles (sampled uniformly at random).

\subsection{Zero-shot target efficacy prediction} \label{sec:zsp}

We evaluate the model's ability to predict transcriptomic responses to gene perturbations without task-specific fine-tuning. This zero-shot perturbation (ZSP) task assesses whether pretrained RNA encoders have implicitly learned gene regulatory relationships that can be accessed via gradient-based inference.

Given a pretrained encoder-decoder pair $(f_{\text{enc}}, f_{\text{dec}})$ and an unperturbed expression profile $\mathbf{x} \in \mathbb{R}^d$ over $d$ genes, the goal is to predict the perturbed expression $\mathbf{x}'$ resulting from up- or down-regulating a target gene $g$, without any perturbation-labeled training data. This setting tests whether the model's learned representations encode sufficient biological structure to extrapolate perturbation effects from observational data alone.

Our approach builds on the gradient flow framework introduced by Bjerregaard et al.~\cite{genes16121439}, who demonstrated that trained single-cell decoders contain perturbation-relevant information accessible via automatic differentiation. By computing gradients of predicted gene expression with respect to latent variables, one obtains vector fields representing infinitesimal changes in expression space. We extend this formalism to bulk RNA encoders and introduce layer-selective perturbation to maximize information content.

\subsubsection{Linear Baseline}

As a baseline, we define a gene-gene interaction matrix $\mathbf{L} \in \mathbb{R}^{d \times d}$ estimated via linear regression from observational data, where $L_{ij}$ represents the expected change in expression of gene $i$ when gene $j$ is perturbed by one unit. Given a perturbation vector $\mathbf{p} \in \mathbb{R}^d$ encoding which genes are perturbed and by how much, the predicted expression is:
\begin{equation}
    \mathbf{x}' = \mathbf{x} + \mathbf{L}\mathbf{p}
\end{equation}
For memory efficiency when $d$ is large, we store a low-rank approximation $\mathbf{L} \approx \mathbf{U}\mathbf{V}^\top$ using truncated SVD, where $\mathbf{U}, \mathbf{V} \in \mathbb{R}^{d \times r}$ and $r \ll d$. We fit the matrix $\mathbf{L}$ on our pretraining human RNA-seq dataset.

\subsubsection{Gradient Flow Perturbation}

We formulate perturbation prediction as gradient-based optimization in the model's representation space. For a set of target genes $\mathcal{G}$ to perturb, we define the objective 

\begin{equation}
    \mathcal{L}_{\text{pert}} = \sum_{g \in \mathcal{G}} \delta_g \cdot \alpha_g \cdot \hat{e}_g
\end{equation}

where $\delta_g \in \{-1, +1\}$ indicates the perturbation direction (inhibition or activation), $\alpha_g > 0$ is the perturbation intensity -- typically 1, and $\hat{e}_g$ is the decoder's predicted expression for gene $g$. Backpropagating this objective yields gradients that indicate how to modify representations to achieve the desired expression change. We implement three perturbation modes.

\paragraph{Latent Space Perturbation.} The encoder maps input expression to embeddings $\mathbf{z} = f_{\text{enc}}(\mathbf{x})$, which are then perturbed along the gradient direction,

\begin{align}
    \mathbf{z}' &= \mathbf{z} + \nabla_{\mathbf{z}} \mathcal{L}_{\text{pert}} \\
    \mathbf{x}' &= f_{\text{dec}}(\mathbf{z}')
\end{align}

Gradients flow only through the decoder during backpropagation, then the perturbed embeddings are decoded to obtain the predicted expression.

\paragraph{Input Space Perturbation.} Alternatively, gradients can propagate through the full encoder-decoder to the input,

\begin{equation}
    \mathbf{x}' = \mathbf{x} \odot \exp(\nabla_{\mathbf{x}} \mathcal{L}_{\text{pert}})
\end{equation}

We use this multiplicative variant to apply perturbations as fold-changes, which preserves non-negativity and provides a natural interpretation in terms of log-fold changes.

\paragraph{Layer-Selective Perturbation.} For transformer encoders with $n$ layers, we perturb at an intermediate layer $\ell$ rather than the final output. Let $\mathbf{h}^{(\ell)}$ denote the hidden states at layer $\ell$:

\begin{align}
    \mathbf{h}^{(\ell)} &= f_{\text{enc}}^{(1:\ell)}(\mathbf{x}) \quad \text{(frozen)} \\
    \mathbf{h}'^{(\ell)} &= \mathbf{h}^{(\ell)} + \nabla_{\mathbf{h}^{(\ell)}} \mathcal{L}_{\text{pert}} \\
    \mathbf{x}' &= f_{\text{dec}}\left(f_{\text{enc}}^{(\ell+1:n)}(\mathbf{h}'^{(\ell)})\right)
\end{align}

We use $\ell = n - 1$ by default, as the final layer representations are specialized for the masked expression reconstruction objective, while earlier layers retain richer gene-level semantic information that better captures regulatory relationships. This aligns with our observation that layer $n-1$ embeddings provide superior representations for downstream fusion tasks (Section~\ref{sec:eva-rna-encoder}).

\subsubsection{Implementation Details}

We normalize gradients to unit $L_2$ norm per sample before applying perturbations, ensuring consistent perturbation magnitude across samples regardless of the objective's scale:
\begin{equation}
    \nabla' = \frac{\nabla}{\|\nabla\|_2 + \epsilon}
\end{equation}
with $\epsilon = 10^{-8}$ for numerical stability. After perturbation, we renormalize expression values to preserve the original library size $\sum_i x_i$, maintaining biologically plausible expression distributions. Both encoder and decoder operate in evaluation mode with frozen parameters throughout inference.

\subsection{Benchmark} \label{sec:benchmark}

To address the absence of standardized evaluation frameworks for biological foundation models in immunology and inflammation, we curated a comprehensive benchmark suite comprising RNA, histology or multimodal-based tasks spanning eight immune-mediated diseases and four tissue types. In contrast to the transformative role that benchmarks such as ImageNet, GLUE, and COCO have played in computer vision and natural language processing~\cite{deng2009imagenet,wang2018glue}, the biological foundation model field currently lacks consensus evaluation protocols, a gap that recent community efforts are beginning to address~\cite{theodoris2024perspectives}. The unique challenges of immunology and inflammation, including heterogeneous patient populations, diverse clinical endpoints, and the need for cross-species translation, motivated the design of domain-specific evaluation tasks that capture clinically relevant prediction scenarios.

Our benchmark follows three guiding principles: (i) \textit{drug development relevance}, with tasks that directly map to translational research decision-making, including cross-species molecular perturbation predictions, clinical treatment response prediction, molecular to clinical severity translation and patient stratification; (ii) \textit{diverse prediction paradigms}, encompassing both supervised learning tasks (classification, regression, perturbation prediction) and zero-shot generalization tasks that test transfer to unseen disease-drug combinations; and (iii) \textit{methodological rigor}, incorporating subject-level data splitting to prevent leakage, evaluation across five random seeds for statistical robustness, and comparison against appropriate baselines (Ridge regression for continuous outcomes, logistic regression for classification, and naive zero-prediction for perturbation tasks). 
\paragraph{RNA-seq benchmark details.}All expression data underwent log$_2$(CPM + 1) normalization, with K-best feature selection applied for linear probe evaluation. Supervised training tasks employed an external test set, or an 80/20 train-test split with 20\% of training data held out for validation. Models were fine-tuned by freezing all weights except last layer. All tasks were run on 5 different seeds, and test results were averaged.

\paragraph{Cross-species evaluation details.} For cross-species tasks (mouse-to-human transfer), BulkRNABert and scGPT require ortholog mapping as these models were trained exclusively on human data. The linear baseline similarly operates on ortholog-mapped features. For fair comparison, we applied the same ortholog mapping to EVA during benchmark evaluation, despite EVA being natively trained on both species. The scaling law analysis (Figure~\ref{fig:scaling-laws}b, Cross-Species treatment effect) demonstrates EVA's capacity to learn cross-species perturbation prediction during training without explicit ortholog mapping, highlighting its ability to discover implicit species alignment through joint pretraining on human and mouse data.

\paragraph{Histology benchmark details.} All details can be found in Section~\ref{apx:histo-eval-method}.

\subsubsection{Discovery - Zero-shot target efficacy prediction}

 The benchmark spans eight diseases and diverse drug mechanism classes, including anti-TNF agents, anti-IL-17/IL-23 antibodies, anti-IL-4/IL-13 biologics, JAK inhibitors, S1P receptor modulators, anti-integrin antibodies, and B-cell targeting therapies.

\begin{table}[htbp]
\centering
\caption{Drugs evaluated in zero-shot perturbation settings across several diseases.}
\label{tab:zsp-coverage}
\begin{tabular}{@{}lc@{}}
\toprule
\tableheaderstyle Disease & Number of drugs \\
\midrule
Atopic dermatitis & 22 \\
Ulcerative colitis & 15 \\
Psoriasis & 12 \\
Crohn's disease & 11 \\
Hidradenitis suppurativa & 11 \\
Psoriatic arthritis & 11 \\
\midrule
\textbf{Total} & \textbf{82}\\
\bottomrule
\end{tabular}
\end{table}

Each couple of disease x drug is evaluated and ranked, and we computed the AUROC of all evaluations.

\subsubsection{Discovery - Gene function prediction}

Beyond sample-level prediction tasks, we evaluate the quality of gene-level representations learned by EVA through their ability to predict gene properties. This evaluation tests whether the model captures biologically meaningful relationships between genes based on their expression patterns, regulatory, and biological contexts.

We designed five complementary tasks to evaluate whether gene embeddings capture functional
relationships relevant to immunology and inflammation. Each task frames gene function prediction
as a multi-label classification problem: given a gene embedding, predict its association with
biological concepts (diseases, pathways, cell types, or Gene Ontology terms \cite{ashburner2000gene}). We train a logistic
regression classifier on frozen gene embeddings and report AUROC averaged across all classes.

\paragraph{Gene-disease association} evaluates the prediction of known therapeutic targets across six immunology and inflammation diseases: alopecia areata, Crohn's disease, psoriasis, rheumatoid arthritis, ulcerative colitis, and hidradenitis suppurativa.

\paragraph{Gene-GO association} evaluates the prediction of Gene Ontology \cite{ashburner2000gene} biological process
annotations across ten immune-related terms, including inflammatory response, innate, and
adaptive immune response, cytokine-mediated signaling, leukocyte migration and activation,
and antigen processing.

\paragraph{Gene-cell type marker association} evaluates prediction of canonical cell type
marker genes across fifteen immune cell populations, including B cells (memory, plasma),
T cell subsets (helper, regulatory, cytotoxic, exhausted, CD4$^+$, CD8$^+$), natural killer
cells, monocytes, macrophages, neutrophils, and mast cells.

\paragraph{Gene-Reactome pathway association} evaluates prediction of Reactome \cite{fabregat2018reactome} pathway membership across twenty expert-curated immune and inflammatory signaling pathways,
including interleukin signaling (IL-1, IL-4, IL-10, IL-17), the NLRP3 inflammasome,
interferon $\alpha/\beta$ and $\gamma$ responses, TLR4 cascade, and NF-$\kappa$B signaling.

\paragraph{Gene-WikiPathways association} evaluates prediction of WikiPathways \cite{kutmon2016wikipathways} membership across eighteen community-curated signaling pathways, including TNF-$\alpha$ signaling,
NOD pathway, B and T cell receptor signaling, and thymic stromal lymphopoietin (TSLP)
signaling, providing an independent pathway knowledge source complementary to Reactome.

For each task, we train logistic regression classifiers on gene embeddings to predict binary associations. We compare EVA gene embeddings against established baselines, including scGPT embeddings (capturing single-cell co-expression patterns), BulkRNABert gene embeddings, and baseline gene embeddings obtained through a PCA on ImmunAtlas RNA-seq data. Evaluation was done through AUPRC with results reported across five random seeds.

\subsubsection{Preclinical - Cross-species treatment effect prediction}

These tasks evaluate whether perturbation patterns learned from mouse models translate to human disease. Models are fine-tuned on mouse datasets and evaluated on human samples as test sets. scGPT and BulkRNABert models weren't trained on mouse data; we used ortholog mapping via NCBI gene identifiers to perform the task. We also used this mapping for EVA for fair comparison between models. This task directly tests the assumption underlying preclinical drug development that molecular responses can be translated across species.
\begin{itemize}
    \item Dupilumab cross-species transfer from mouse lung, affected by asthma to human skin affected by atopic dermatitis, testing conservation of IL-4R$\alpha$ blockade signatures across tissues, species and conditions;
    \item TNF-$\alpha$ inhibitor cross-species transfer in synovial joints affected by rheumatoid arthritis; 
\end{itemize}

A preprocessing step selects the top 1,200 most variable genes per dataset across disease, treated and control samples, to focus the evaluation on biologically meaningful changes. Evaluation metrics include mean absolute error (MAE), relative MAE normalized to baseline expression (RMAE), and Pearson and Spearman correlations computed on expression \textit{changes} per sample (post-treatment minus baseline) rather than absolute values. All metrics are compared against a naive zero-prediction baseline that assumes no treatment effect.

\subsubsection{Preclinical - Molecular perturbations prediction}

Molecular perturbation prediction tasks assess the model's ability to predict full transcriptomic changes following therapeutic intervention, rather than scalar clinical outcomes. Given baseline gene expression profiles, the model must predict post-treatment expression values for each gene, a substantially more challenging task that requires modeling the complex regulatory consequences of pharmacological perturbation.

We curated five tasks that perturbation capabilities across different diseases and species. In particular, two cross-disease tasks evaluate whether perturbation responses generalize across indications sharing the same drug target: we train the model on samples from one disease and evaluate on another disease, where patients were treated with the same therapy. This benchmark tests whether the model has learned common and transferable signatures for the same mechanism-of-action across diseases. Tasks are composed of

\begin{itemize}
    \item Rituximab response in Sjögren's Disease, evaluating B-cell depletion signatures within species;
    \item  TNF-KO perturbations in a mouse colitis model;
    \item bidirectional adalimumab cross-disease transfer between Hidradenitis Suppurativa and Psoriasis, testing whether anti-TNF signatures generalize across cutaneous inflammatory conditions
\end{itemize}

Evaluation and preprocessing were identical to cross-species treatment effect. A summary of all perturbation tasks is presented in Table~\ref{tab:perturbation-tasks}. 

\begin{table}[htbp]
\centering
\caption{Perturbation prediction tasks evaluating cross-species and cross-disease transfer. Conditions after arrows indicate test set data. HS: hidradenitis suppurativa; Pso: psoriasis; RA: rheumatoid arthritis; SjD: Sjögren's Disease.}
\label{tab:perturbation-tasks}
\begin{tabular}{@{}lllllll@{}}
\toprule
\tableheaderstyle Drug & Drug Target & Disease & Tissue & Species\\
\midrule
Dupilumab & Anti-IL-4R$\alpha$ & Asthma$\rightarrow$Atopic Dermatitis & Lung$\rightarrow$Skin & Mouse$\rightarrow$Human \\
TNFi & Anti-TNF & RA & Synovial Joint & Mouse$\rightarrow$Human \\
Rituximab & Anti-CD20 & SjD & Blood & Human \\
Rituximab & Anti-CD20 & SjD & Salivary gland & Human \\
TNF-$\alpha$ KO & Anti-TNF & IBDs & GI & Mouse \\
Adalimumab & Anti-TNF & HS$\rightarrow$Pso & Skin & Human \\
Adalimumab & Anti-TNF & Pso$\rightarrow$HS & Skin & Human \\
\bottomrule
\end{tabular}
\end{table}
\subsubsection{Clinical - Treatment outcome prediction}

Six tasks evaluate binary prediction of therapeutic response in inflammatory bowel disease (IBD). We predict one type of outcomes: endoscopic remission.
For endoscopic remission, we assess three anti-TNF agents (Infliximab, Adalimumab, and Vedolizumab) in colon and ileum biopsies. This includes cross-treatment generalization tasks where models trained on one anti-TNF agent (Infliximab or Adalimumab) are tested on the other, evaluating whether response signatures generalize across similar mechanisms of action. We also include cross-modality tasks with Vedolizumab, where models are trained on one transcriptomic platform (RNA-seq or microarray) and tested on the other, assessing robustness to technical variation.

\begin{table}[htbp]
\centering
\caption{Treatment response prediction tasks. IBD: inflammatory bowel disease. Treatment or technology after arrow are used in the test set.}
\label{tab:classification-tasks}
\begin{tabular}{@{}llll@{}}
\toprule
\tableheaderstyle Disease & Tissue & Drug/Context & Target \\
\midrule
IBD & Colon/Ileum & Infliximab & Endoscopic Remission \\
IBD & Colon/Ileum & Adalimumab & Endoscopic Remission \\
IBD & Colon/Ileum & Adalimumab $\rightarrow$ Infliximab & Endoscopic Remission \\
IBD & Colon/Ileum & Infliximab $\rightarrow$ Adalimumab  & Endoscopic Remission \\
IBD & Colon (microarray $\rightarrow$ RNA-seq) & Vedolizumab & Endoscopic Remission \\
IBD & Colon (RNA-seq $\rightarrow$ microarray) & Vedolizumab & Endoscopic Remission \\
\bottomrule
\end{tabular}
\end{table}

\subsubsection{Clinical - Disease activity assessment from molecular state}

Predicting clinical outcomes from molecular changes is notoriously challenging in I\&I. We evaluate our model capabilities to bridge this gap with 14 tasks predicting clinical severity indices from gene expression profiles. These scores span three assessment modalities, clinical examination, endoscopy, and histopathology, testing whether transcriptomic signatures capture the full continuum of disease severity rather than merely categorical distinctions.

\paragraph{Atopic dermatitis.} We predict two complementary severity measures: the Eczema Area and Severity Index (EASI), a composite score integrating lesion extent and intensity across body regions (continuous scale 0--72) that serves as the primary endpoint in most clinical trials; and SCORAD (Scoring Atopic Dermatitis), which additionally incorporates subjective symptoms including pruritus and sleep disturbance (continuous scale 0--103).

\paragraph{Crohn's disease.} Three tasks capture distinct dimensions of disease activity. The Global Histologic Activity Score (GHAS) quantifies histological inflammation severity from ileal and colonic biopsies, reflecting the current treatment target of mucosal healing. The Harvey-Bradshaw Index (HBI) provides a rapid clinical assessment based on general well-being, abdominal pain, and stool frequency. The Simple Endoscopic Score for Crohn's Disease (SES-CD) integrates endoscopic findings, including ulcer size, ulcerated surface area, affected surface area, and presence of stenosis.

\paragraph{Psoriasis.} The Psoriasis Area and Severity Index (PASI) represents the gold-standard assessment, combining affected body surface area with erythema, induration, and scaling severity (continuous scale 0--72). PASI improvement thresholds (PASI75, PASI90, PASI100) define categorical response criteria in clinical trials.

\paragraph{Rheumatoid arthritis.} Two joint-level assessments capture inflammatory activity: the Swollen Joint Count (SJC28), an objective measure of active synovitis across 28 joints, and the Tender Joint Count (TJC28), reflecting patient-reported joint pain. Both are components of the ACR (American College of Rheumatology) response criteria.

\paragraph{Sjögren's Disease.} Three tasks address systemic disease activity. The EULAR Sjögren's Disease Disease Activity Index biological domain (ESSDAI-BIO) captures systemic manifestations requiring immunosuppressive treatment. Immunoglobulin levels (IgA and IgG) serve as serological markers of B-cell hyperactivity, with hypergammaglobulinemia being a hallmark of this condition.

\paragraph{Ulcerative colitis.} Three complementary assessments span different modalities. The Mayo endoscopic subscore (0--3 scale) quantifies mucosal inflammation severity, with endoscopic remission (Mayo 0--1) representing a key treatment target. The Nancy histological index provides a validated histological assessment that predicts long-term clinical outcomes. The Simple Clinical Colitis Activity Index (SCCAI) offers a symptom-based measure of disease activity.

\begin{table}[htbp]
\centering
\caption{Disease severity assessment tasks. AD: atopic dermatitis; CD: Crohn's disease; GI: gastrointestinal; Pso: psoriasis; RA: rheumatoid arthritis; SjD: Sjögren's Disease; UC: ulcerative colitis.}
\label{tab:regression-tasks}
\begin{tabular}{@{}lllll@{}}
\toprule
\tableheaderstyle Disease & Clinical Score & Tissue & Score Type \\ 
\midrule
AD & EASI & Skin & Composite\\
AD & SCORAD & Skin & Composite\\
CD & GHAS & GI tract & Histological \\
CD & HBI & GI tract & Clinical \\
CD & SES-CD & GI tract & Endoscopic \\
Pso & PASI & Skin & Composite \\
RA & SJC28 & Synovial & Clinical \\
RA & TJC28 & Synovial & Clinical \\
SjD & ESSDAI-BIO & Blood & Clinical \\
SjD & IgA & Blood & Serological \\
SjD & IgG & Blood & Serological \\
UC & Mayo endoscopic & GI tract & Endoscopic \\
UC & Nancy Index & GI tract & Histological \\
UC & SCCAI & GI tract & Clinical \\
\bottomrule
\end{tabular}
\end{table}

\subsubsection{Clinical - Endotype classification}

Two tasks address the classification of rheumatoid arthritis endotypes based on synovial tissue architecture, distinguishing lymphoid, myeloid, and fibroid phenotypes that associate with distinct disease trajectories and treatment responses. We evaluate this task from both peripheral blood (testing the feasibility of non-invasive pathotype assignment) and synovial joint tissue (the gold-standard classification).

Results were evaluated using AUROC.

\subsubsection{Clinical - Histological diagnosis} \label{histology-tasks-presentation}

\paragraph{Sjögren's Disease.} Two binary classification tasks assess autoimmune salivary gland pathology. The focus score task distinguishes patients with significant lymphocytic infiltration (focus score $\geq 1$, the diagnostic threshold for focal sialadenitis) from those without. The diagnosis task classifies biopsies as Sjögren's Disease positive or negative based on clinical diagnosis, testing whether histological patterns alone capture the full diagnostic picture.

\subsubsection{Clinical - Histological scoring} 

\paragraph{Inflammatory bowel disease.} Two histological scoring tasks evaluate disease activity in IBD using the IBDome dataset~\cite{Plattner2025}. The normalized modified Naini-Cortina score quantifies chronic inflammatory changes in ileal and colonic biopsies, while the normalized modified Riley score assesses acute inflammation severity, with histological remission increasingly recognized as a treatment target beyond endoscopic healing. Both tasks are formulated as regression problems predicting continuous scores from whole-slide images.

\subsection{Latent space interpretability with sparse auto-encoders} \label{method:sae}

Recent work in mechanistic interpretability showed that sparse dictionary learning methods can decompose the internal representations of deep learning models into a sparse combination of interpretable concepts. Non-Negative Matrix Factorization has shown promising results in image and text classification models \cite{fel2023craft, jourdan2023cockatiel}, and Sparse Auto-Encoders have extracted millions of concepts in generative language models \cite{bricken2023towards}. Several works have adapted these approaches to biological foundation models, a few of them focusing on transcriptomic models \cite{Schuster2024CanSA, claye2025discovering, claye2025framework, pedrocchi2025sparse}. Following \cite{claye2025discovering}, we extracted concepts in sample-level embeddings and used attribution methods to further identify the genes characterizing each concept.

\paragraph{TopK Sparse Auto-Encoders (SAEs).}

Following the results in \cite{fel2025archetypal}, we use SAEs, which achieve better reconstructions at a fixed sparsity level. We also chose to use topK-SAEs following the work of \cite{gaoscaling}, which simplifies the tuning and improves the reconstruction-sparsity frontier compared to vanilla SAEs.

Given a sample embedding $\mathbf{a} \in \mathbb{R}^d$ from the model, topK-SAE first computes the corresponding concept activations $\mathbf{u} \in \mathbb{R}^c$ with an encoder $\mathbf{u} = \text{ReLU}(\text{TopK}(\mathbf{a}\mathbf{W}_e+\mathbf{b}_e))$. Given the concept activations, the decoder reconstructs the sample embedding as $\mathbf{a'} = \mathbf{u}\mathbf{W}_d$, where each vector in $\mathbf{W}_d$ corresponds to an interpretable direction in the latent space. The auto-encoder is trained on a dataset of embeddings described in the next paragraph, with a reconstruction loss between $\mathbf{a}$ and $\mathbf{a}'$.

\paragraph{Dataset of sample-level embeddings.} To train the topK-SAE model, we form a dataset of $n$ embeddings. We chose to extract concepts from the last CLS token representations to capture sample-level concepts. We used all samples from the pretraining dataset, except that we randomly selected 40,000 microarray samples to balance the training dataset. For each sample, we generated 3 embeddings corresponding to 3 random selections of 2000 genes. The final training dataset is composed of 442,332 embeddings of dimension $d=768$.

\begin{figure}[!ht]
    \centering
    \includegraphics[width=\textwidth]{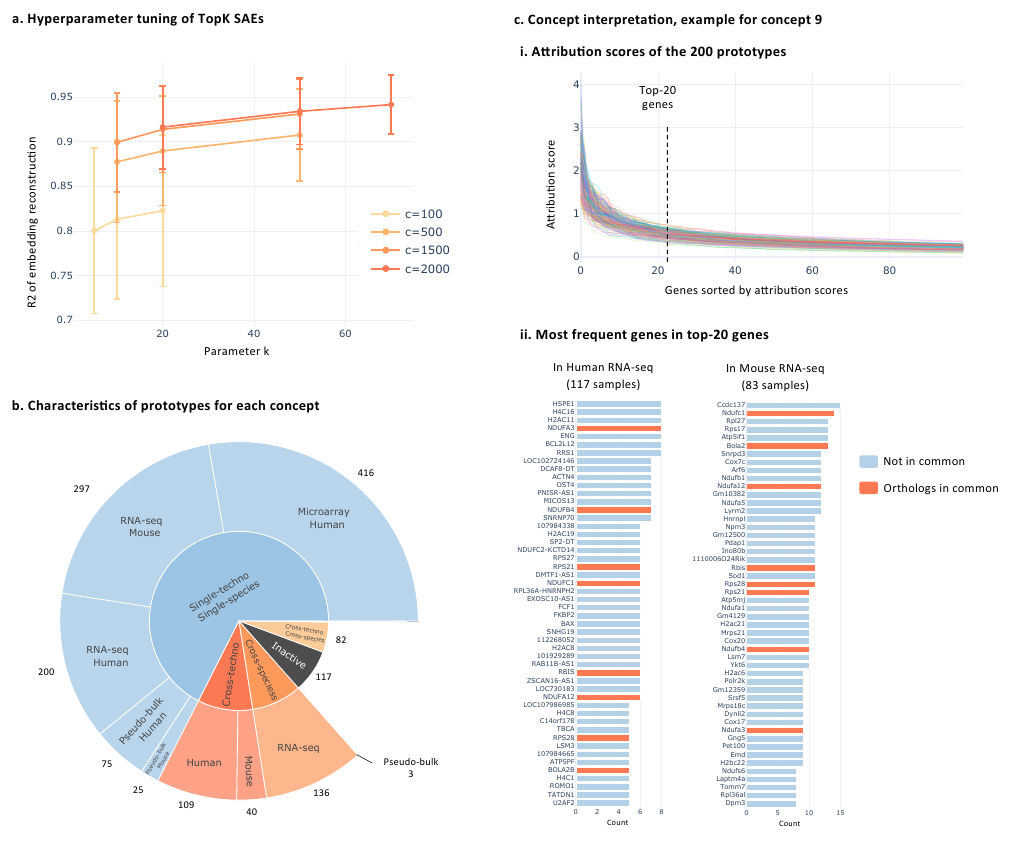}
    \caption{Latent space interpretability with Sparse Auto-encoders. a. Hyperparameter search for topK-SAEs for concept extraction in contextualized sample representations (last CLS token). Embedding reconstruction error at different $c$ and $k$. A score of 1 means that the embeddings are reconstructed perfectly from the concept activations, while a score of 0 means that the reconstruction is no better than the mean. b. Distribution of concepts in terms of species and technologies represented in the 200 samples with the highest concept activation ("prototypes"). c. Example of attribution results for concept 9. (a) Attribution scores from the highest score to the lowest for each prototype. For all prototypes, the top 20 genes have a high attribution score. (b) Most frequent genes in the top-20 genes of each prototype, by species. Given the mapping of orthologous genes between human and mouse, several genes that are most frequently important in human samples are also important in mouse samples, suggesting a shared interpretation across the species.}
    \label{fig:panel_sae}
\end{figure}

\paragraph{Hyperparameter search.} We trained several topK-SAEs at different numbers of concepts $c$ and different sparsity constraints with $k$. We monitored the embedding reconstruction error with the $R^2$, as shown in Figure~\ref{fig:panel_sae}a. We selected $c=1500$ and $k=20$, which gives low reconstruction error while keeping the number of concepts moderate.

\paragraph{Cross-technologies and cross-species concepts.} To detect concepts that are shared across modalities or species, we look at the technologies and species among the 200 samples that maximally activate each concept. We annotate with the technology or specie if at least 20 samples from a technology or a species are among the 200 prototypes. 

\paragraph{Concept interpretation.} To interpret a concept biologically, we computed attribution scores for 200 samples that highly activate the concept (referred to as a prototype), the scores indicate how each gene contributed to the activation of the concept. We used Integrated Gradients \cite{sundararajan2017axiomatic} with the Captum implementation \cite{kokhlikyan2020captum}, setting the hyperparameters to 20 steps and a baseline of 0, corresponding to an unexpressed gene. After looking at a few attribution scores (Example in Figure~\ref{fig:panel_sae}c.i), we decided to focus on the top 20 genes in each prototype. We visualized the most frequent genes in the top 20, for each prototype, separated by technologies and species. An example for concept 9 is given in Figure~\ref{fig:panel_sae}c.ii. We identified several concepts with important genes shared across technologies or species (as determined by orthologs mapping, example in Figure~\ref{fig:panel_sae}c.ii), along with a shared biological interpretation.

\section{Discussion}

In this work, we introduced EVA, the first cross-species, multimodal foundation model specifically designed for immunology and inflammation. Our results demonstrate that EVA successfully integrates heterogeneous data sources, spanning technologies (microarray, bulk RNA-seq, pseudobulked single-cell), species (human and mouse), and modalities (transcriptomics and histology), into unified patient-level representations that capture biologically meaningful signals across the drug discovery pipeline.

\paragraph{Cross-species and cross-technology integration addresses key translational barriers.}
A fundamental challenge in drug development is the translation of findings from preclinical models to human diseases. Our analysis reveals that EVA progressively aligns mouse genes with their human orthologs during training, with immune genes achieving particularly strong cross-species alignment. The sparse autoencoder analysis further identified interpretable concepts that detect shared and coherent biological programs across species and technologies, such as lymphocyte immune programs and epithelial differentiation signatures. These findings suggest that EVA captures conserved molecular mechanisms that underlie immune-mediated diseases, potentially enabling more reliable cross-species predictions for therapeutic development. The ability to integrate legacy microarray datasets alongside modern RNA-seq data is particularly valuable, as it unlocks decades of accumulated transcriptomic data that have been historically difficult to combine due to batch effects.

\paragraph{Scaling laws provide a roadmap for continued improvement.}
A notable finding is that EVA exhibits clear, predictable scaling behavior, a property that has been inconsistently observed in biological foundation models. In contrast to previous reports suggesting diminishing returns beyond 100M parameters in gene expression models \cite{ho2024scaling}, EVA-RNA demonstrates continued improvement up to 300M parameters with no sign of plateauing. Critically, we show that pretraining loss improvements translate reliably to downstream task performance across diverse evaluation categories, addressing a key concern raised in recent foundation model benchmarks \cite{kedzierska2025zero}. This predictable relationship between compute investment and model capability provides a principled basis for future scaling decisions and suggests that larger EVA models may yield further improvements.

\paragraph{Benchmark design reflects drug development priorities.}
The I\&I benchmark we introduce encompasses tasks directly relevant to translational research and drug development, such as target-disease association for discovery, cross-species perturbation prediction for preclinical development, and patient stratification with treatment response prediction for biomarkers discovery and clinical development. By evaluating models on tasks that map to actual decision points in drug development, rather than purely technical metrics, we aim to accelerate the adoption of foundation models in preclinical, translational and clinical research. The zero-shot perturbation tasks are particularly noteworthy, as they test whether models can generalize to novel disease-drug combinations without task-specific training, a scenario that closely mirrors real-world drug development, where historical data for new indications is often scarce or unavailable.

\paragraph{Mechanistic interpretability.}
The interpretability analysis using sparse autoencoders represents a promising direction for understanding what biological knowledge foundation models encode. While we identified several interpretable concepts, systematic characterization of the full latent space remains an open challenge. Developing automated methods to annotate and validate discovered concepts against known biology could enhance trust in model predictions and potentially reveal novel biological insights.

\paragraph{Limitations and future directions.}
Several limitations of the current work suggest directions for future research. First, while EVA integrates transcriptomics and histology, other data modalities central to drug discovery---including proteomics, metabolomics, and spatial transcriptomics---remain unincorporated. Extending EVA to these modalities could capture additional layers of biological regulation relevant for therapeutic response. Second, our histology training data, while diverse within I\&I, remain smaller than datasets available in oncology; continued curation of I\&I pathology datasets will be essential for improving EVA-H. Third, although EVA demonstrates strong performance on perturbation prediction tasks, the model currently operates on bulk or pseudobulk representations, potentially obscuring cell-type-specific drug responses that may be critical for disentangling cell-level contributions and understanding therapeutic mechanisms. Future versions could incorporate cell-type deconvolution or operate directly on single-cell data while maintaining sample-level coherence. Finally, prospective validation of EVA's predictions in clinical settings will be essential to establish its utility for drug development decision-making. The treatment response prediction tasks evaluated here use retrospective data; demonstrating that EVA can improve patient selection or predict outcomes in ongoing clinical trials would provide compelling evidence of the translational potential and impact.

\paragraph{Broader implications for biological foundation models.}
Our work suggests several lessons for the broader foundation model community. First, domain specificity may be advantageous: by focusing on immunology and inflammatory diseases, EVA can leverage the shared pathogenic mechanisms across conditions in this therapeutic area, potentially enabling more effective transfer learning than general-purpose biological models. Second, evaluation frameworks should be aligned with intended applications; the disconnect between pretraining objectives, the low-level metrics, and downstream utility observed in some foundation models may in part reflect benchmark design that does not capture relevant biological tasks. Third, cross-species training, often overlooked in favor of human-only datasets, may be particularly valuable for applications in drug development where preclinical translation is a major bottleneck.

\section{Conclusion}

EVA, our multimodal foundation model, integrates transcriptomic and histology data across species and technologies to produce unified patient-level representations for immunology and inflammation research. EVA demonstrates clear scaling laws, with pretraining improvements translating consistently to downstream task performance across a comprehensive benchmark spanning drug discovery, preclinical translation, and clinical applications. Through sparse autoencoder analysis, we identified interpretable features that reveal how EVA learns shared biological concepts across species and data modalities. By releasing an open version of EVA-RNA, we aim to accelerate translational research in immune-mediated diseases and provide the community with a foundation to develop more effective therapies for conditions affecting hundreds of millions of patients worldwide.

\section*{Acknowledgments}
This project was partially supported by computational and storage resources from the GENCI at IDRIS, thanks to the grants 2025-AD010316784 and 2025-AD010316294 on the supercomputer Jean Zay's A100 and H100 partitions. We are grateful to Andrei Zinovyev for his thoughtful and constructive feedback on this manuscript.

\bibliographystyle{unsrt}  
\bibliography{biblio}

\newpage

\section{Appendix}

\subsection{Results per task}\label{apx:detailed-results}
We developed a comprehensive benchmark to evaluate whether EVA learns transferable representations across the drug development pipeline, from target discovery through preclinical translation to clinical applications (Section~\ref{sec:benchmark}). We assess both unimodal encoders -- EVA-RNA for transcriptomics and EVA-H for histology -- against state-of-the-art biological foundation models and statistical baselines.

EVA-RNA is a transformer encoder pretrained on ImmunAtlas, our curated corpus of 545k transcriptomic samples spanning human and mouse, three technologies, and over 50 immunological conditions (Sections~\ref{sec:rna-datasets} and~\ref{sec:eva-rna-encoder}). We trained three model sizes (7M, 60M, and 300M parameters) to investigate scaling behavior, and compared against scGPT, BulkRNABert, and statistical baselines. EVA-H is a vision transformer trained on 4k whole-slide images from I\&I-relevant tissues (Section~\ref{sec:eva-h}), benchmarked against Hibou-B/L, and CHIEF (Section \ref{apx:histo-eval-method}).

Tables~\ref{tab:benchmark-unimodal} and~\ref{tab:histo_results_aggregated} summarize performance across task categories. EVA-RNA achieves state-of-the-art results on six of seven task categories, with consistent improvement from 7M to 300M parameters. The largest gains over competing foundation models appear in zero-shot perturbation (0.693 vs.\ 0.539 for scGPT) and treatment outcome prediction (0.736 vs.\ 0.581), suggesting that patient-level pretraining on bulk I\&I data captures clinically relevant signatures that single-cell only models trained on general broader data miss. For histology, EVA-H achieves competitive performance, ranking first or second on both Sjögren's disease activity and IBD histological scoring. EVA-RNA detailed results can be found in Table \ref{tab:benchmark-unimodal-detailed}. EVA-H detailed results can be found in Table \ref{tab:histo-detailed-results}.

% Old version
%\input{table_eva_r_results}

% This new table (same content) makes this big Table stay in its section for easier reading
% Requires: \usepackage{array}, \usepackage{caption}
\begin{center}
\begin{minipage}{\textwidth}
    \centering
    \scriptsize
    \captionof{table}{EVA-RNA performance on all I\&I benchmark tasks (mean $\pm$ std). Metrics: AUROC for Zero-Shot target efficacy, Stratification into endotype, Clinical treatment outcome; AUPRC for Gene function; Pearson for Molecular Perturbation, Cross-Species treatment effect, Molecular to clinical activity. Bold/underline: best/second-best. Arrows: transfer direction.}
    \label{tab:benchmark-unimodal-detailed}
    \begin{tabular}{@{}>{\raggedright\arraybackslash}p{5cm}cccccc@{}}
        \toprule
        Task & 7M & 60M & 300M & scGPT & BulkRNABert & Baseline \\
        \midrule
        \multicolumn{7}{l}{\textbf{Zero-shot target efficacy (AUROC)}} \\
        \midrule
        Zero-Shot Perturbation & \underline{0.66} & 0.62 & \textbf{0.69} & 0.54 & -- & 0.57 \\
        \midrule
        \multicolumn{7}{l}{\textbf{Gene function (AUPRC)}} \\
        \midrule
        CellType & 0.37 $\pm$ 0.23 & \underline{0.45 $\pm$ 0.22}& 0.41 $\pm$ 0.19 & \textbf{0.48 $\pm$ 0.24} & 0.34 $\pm$ 0.21 & 0.39 $\pm$ 0.23\\
        Disease & \underline{0.33 $\pm$ 0.31} & 0.31 $\pm$ 0.36 & \textbf{0.46 $\pm$ 0.34} & 0.24 $\pm$ 0.32 & 0.22 $\pm$ 0.26 & 0.25 $\pm$ 0.30 \\
        GO & 0.40 $\pm$ 0.29 & \underline{0.43 $\pm$ 0.30} & \textbf{0.47 $\pm$ 0.27} & 0.34 $\pm$ 0.28 & 0.30 $\pm$ 0.22 & 0.29 $\pm$ 0.28 \\
        Reactome & 0.50 $\pm$ 0.29 & \underline{0.58 $\pm$ 0.27} & \textbf{0.64 $\pm$ 0.29} & 0.36 $\pm$ 0.22 & 0.30 $\pm$ 0.23 & 0.32 $\pm$ 0.25 \\
        WikiPathways & 0.33 $\pm$ 0.22 & \underline{0.42 $\pm$ 0.24} & \textbf{0.49 $\pm$ 0.27} & 0.37 $\pm$ 0.21 & 0.27 $\pm$ 0.20 & 0.38 $\pm$ 0.23 \\
        \midrule
        \multicolumn{7}{l}{\textbf{Molecular perturbation (Pearson)}} \\
        \midrule
        Anti-TNF (IBD mice) & 0.54 $\pm$ 0.13 & \underline{0.66 $\pm$ 0.17} & \textbf{0.70 $\pm$ 0.16} & 0.37 $\pm$ 0.19 & 0.49 $\pm$ 0.18 & 0.59 $\pm$ 0.21 \\
        Adalimumab (HS $\rightarrow$ Pso) & \textbf{0.48 $\pm$ 0.01} & \underline{0.46 $\pm$ 0.01} & 0.45 $\pm$ 0.01 & 0.41 $\pm$ 0.02 & 0.41 $\pm$ 0.03 & -0.19 $\pm$ 0.02 \\
        Adalimumab (Pso $\rightarrow$ HS) & 0.27 $\pm$ 0.04 & \underline{0.29 $\pm$ 0.01} & \textbf{0.32 $\pm$ 0.00} & 0.23 $\pm$ 0.01 & 0.24 $\pm$ 0.02 & -0.13 $\pm$ 0.00 \\
        Rituximab (SjD blood) & 0.77 $\pm$ 0.02 & \textbf{0.78 $\pm$ 0.03} & \underline{0.77 $\pm$ 0.02} & 0.75 $\pm$ 0.04 & 0.74 $\pm$ 0.02 & 0.60 $\pm$ 0.04 \\
        Rituximab (SjD salivary) & 0.50 $\pm$ 0.11 & \textbf{0.52 $\pm$ 0.10} & 0.50 $\pm$ 0.10 & \underline{0.51 $\pm$ 0.10} & 0.49 $\pm$ 0.12 & -0.01 $\pm$ 0.09 \\
        \midrule
        \multicolumn{7}{l}{\textbf{Cross-species treatment effect (Pearson)}} \\
        \midrule
        Dupilumab & 0.47 $\pm$ 0.00 & \underline{0.48 $\pm$ 0.01} & 0.48 $\pm$ 0.00 & 0.47 $\pm$ 0.00 & \textbf{0.48 $\pm$ 0.00} & 0.05 $\pm$ 0.00 \\
        TNFi RA & 0.41 $\pm$ 0.00 & \textbf{0.41 $\pm$ 0.00} & \underline{0.41 $\pm$ 0.00} & 0.41 $\pm$ 0.00 & 0.39 $\pm$ 0.00 & 0.00 $\pm$ 0.00 \\
        \midrule
        \multicolumn{7}{l}{\textbf{Molecular to clinical activity (Pearson)}} \\
        \midrule
        Blood IgA & 0.37 $\pm$ 0.11 & \textbf{0.39 $\pm$ 0.09} & 0.37 $\pm$ 0.11 & 0.28 $\pm$ 0.13 & 0.18 $\pm$ 0.12 & \underline{0.39 $\pm$ 0.10} \\
        Blood IgG & 0.47 $\pm$ 0.07 & 0.51 $\pm$ 0.07 & \underline{0.51 $\pm$ 0.10} & 0.41 $\pm$ 0.06 & 0.25 $\pm$ 0.15 & \textbf{0.54 $\pm$ 0.09} \\
        Digestive GHAS7 & 0.60 $\pm$ 0.05 & \underline{0.61 $\pm$ 0.04} & 0.57 $\pm$ 0.02 & 0.58 $\pm$ 0.05 & 0.55 $\pm$ 0.07 & \textbf{0.61 $\pm$ 0.03} \\
        Digestive SES-CD & 0.39 $\pm$ 0.07 & \underline{0.45 $\pm$ 0.10} & \textbf{0.46 $\pm$ 0.09} & 0.39 $\pm$ 0.07 & 0.35 $\pm$ 0.07 & 0.44 $\pm$ 0.08 \\
        ESSDAI Bio & 0.35 $\pm$ 0.10 & 0.40 $\pm$ 0.08 & \textbf{0.41 $\pm$ 0.16} & 0.33 $\pm$ 0.08 & 0.13 $\pm$ 0.15 & \underline{0.40 $\pm$ 0.08} \\
        Endoscopic Mayo & 0.64 $\pm$ 0.13 & \textbf{0.67 $\pm$ 0.13} & 0.64 $\pm$ 0.15 & 0.65 $\pm$ 0.15 & 0.60 $\pm$ 0.15 & \underline{0.66 $\pm$ 0.12} \\
        HBI & 0.14 $\pm$ 0.05 & 0.15 $\pm$ 0.08 & \textbf{0.24 $\pm$ 0.07} & 0.14 $\pm$ 0.06 & 0.12 $\pm$ 0.08 & \underline{0.19 $\pm$ 0.06} \\
        Nancy Histological Index & \underline{0.76 $\pm$ 0.11} & 0.75 $\pm$ 0.07 & 0.73 $\pm$ 0.09 & 0.75 $\pm$ 0.09 & 0.71 $\pm$ 0.14 & \textbf{0.76 $\pm$ 0.12} \\
        RA SJC28 & 0.27 $\pm$ 0.22 & \underline{0.36 $\pm$ 0.21} & \textbf{0.44 $\pm$ 0.16} & 0.29 $\pm$ 0.20 & 0.24 $\pm$ 0.18 & 0.36 $\pm$ 0.20 \\
        RA TJC28 & 0.23 $\pm$ 0.24 & 0.21 $\pm$ 0.21 & \underline{0.27 $\pm$ 0.17} & 0.24 $\pm$ 0.20 & 0.19 $\pm$ 0.21 & \textbf{0.28 $\pm$ 0.25} \\
        SCCAI & 0.52 $\pm$ 0.04 & \textbf{0.61 $\pm$ 0.05} & \underline{0.59 $\pm$ 0.07} & 0.50 $\pm$ 0.06 & 0.49 $\pm$ 0.09 & 0.57 $\pm$ 0.05 \\
        Skin EASI & \textbf{0.32 $\pm$ 0.19} & 0.28 $\pm$ 0.16 & 0.29 $\pm$ 0.20 & 0.29 $\pm$ 0.16 & 0.24 $\pm$ 0.15 & \underline{0.30 $\pm$ 0.22} \\
        Skin PASI & 0.18 $\pm$ 0.15 & 0.17 $\pm$ 0.22 & \textbf{0.26 $\pm$ 0.24} & 0.10 $\pm$ 0.16 & 0.21 $\pm$ 0.08 & \underline{0.25 $\pm$ 0.22} \\
        Skin SCORAD & 0.11 $\pm$ 0.16 & \underline{0.21 $\pm$ 0.11} & \textbf{0.27 $\pm$ 0.14} & 0.07 $\pm$ 0.13 & 0.09 $\pm$ 0.11 & 0.21 $\pm$ 0.19 \\
        \midrule
        \multicolumn{7}{l}{\textbf{Stratification into endotypes (AUROC)}} \\
        \midrule
        RA blood & 0.64 $\pm$ 0.06 & \textbf{0.66 $\pm$ 0.12} & \underline{0.65 $\pm$ 0.12} & 0.52 $\pm$ 0.07 & 0.58 $\pm$ 0.08 & 0.63 $\pm$ 0.12 \\
        RA joint & 0.90 $\pm$ 0.02 & \underline{0.91 $\pm$ 0.02} & \textbf{0.92 $\pm$ 0.01} & 0.89 $\pm$ 0.02 & 0.81 $\pm$ 0.03 & 0.90 $\pm$ 0.01 \\
        \midrule
        \multicolumn{7}{l}{\textbf{Clinical treatment outcome (Endoscopic remission) (AUROC)}} \\
        \midrule
        IBD, adalimumab (ADA) & 0.46 $\pm$ 0.07 & \underline{0.56 $\pm$ 0.13} & \textbf{0.76 $\pm$ 0.07} & 0.40 $\pm$ 0.12 & 0.37 $\pm$ 0.16 & 0.48 $\pm$ 0.17 \\
        IBD, infliximab (IFX) & 0.53 $\pm$ 0.19 & 0.55 $\pm$ 0.29 & \textbf{0.80 $\pm$ 0.24} & 0.50 $\pm$ 0.13 & 0.48 $\pm$ 0.18 & \underline{0.62 $\pm$ 0.10} \\
        IBD, ADA $\rightarrow$ IFX & \underline{0.65 $\pm$ 0.08} & 0.56 $\pm$ 0.08 & 0.50 $\pm$ 0.05 & 0.59 $\pm$ 0.05 & 0.64 $\pm$ 0.07 & \textbf{0.66 $\pm$ 0.04} \\
        IBD, IFX $\rightarrow$ ADA & \textbf{0.63 $\pm$ 0.02} & 0.57 $\pm$ 0.06 & 0.57 $\pm$ 0.04 & 0.61 $\pm$ 0.04 & 0.57 $\pm$ 0.02 & \underline{0.62 $\pm$ 0.02} \\
        IBD, vedolizumab, RNAseq $\rightarrow$ microarray  & 0.52 $\pm$ 0.11 & \underline{0.62 $\pm$ 0.03} & \textbf{0.65 $\pm$ 0.04} & 0.47 $\pm$ 0.09 & 0.47 $\pm$ 0.07 & 0.61 $\pm$ 0.01 \\
        IBD, vedolizumab, microarray $\rightarrow$ RNAseq & \textbf{0.64 $\pm$ 0.02} & 0.60 $\pm$ 0.05 & 0.63 $\pm$ 0.05 & 0.37 $\pm$ 0.03 & 0.44 $\pm$ 0.07 & \underline{0.64 $\pm$ 0.02} \\
        \bottomrule
    \end{tabular}
\end{minipage}
\end{center}

\subsection{WSIs preprocessing details}\label{apx:wsi-preprocessing}

Whole slide images (WSIs) are preprocessed using a multi-stage tissue segmentation and tile extraction pipeline. First, tissue regions are automatically segmented from the background by converting the downsampled WSIs to HSV color space and extracting the saturation channel, which is then processed with median blurring (kernel size 5) followed by adaptive mean thresholding. Morphological closing operations (kernel size 9) are applied to fill small gaps and smooth contour boundaries. Tissue contours and internal holes (cavities within tissue) are identified using OpenCV's hierarchical contour detection with RETR\_CCOMP mode. Contours are filtered based on area thresholds relative to a reference tile size of $512 \times 512$ pixels: tissue regions must exceed 16 reference tiles in area, while holes larger than 4 reference tiles are preserved (up to 8 holes per tissue region) to avoid extracting tiles from artifacts or damaged tissue areas. Within each valid tissue contour, tiles of size $224 \times 224$ pixels are extracted using a sliding window with step size matching the tile size (non-overlapping grid). Tile inclusion is determined by the ``single corner in contour'' criterion, which accepts a tile if at least one corner point (defined by a center shift parameter of 0.5) falls within the tissue boundary and outside any holes. Optionally, tiles are filtered to exclude predominantly white regions (HSV saturation mean $<$ 15) or black regions (RGB mean $<$ 50 per channel), which typically correspond to glass background or marker artifacts. Tile coordinates and optional image data are stored in HDF5 format for efficient batch loading during training. All processing is performed at 20$\times$ magnification, with automatic rescaling when slide's native magnification differs from the target magnification.

\subsection{Histology evaluation framework} \label{apx:histo-eval-method}
This section details how we evaluated histology encoders. Table~\ref{tab:histo-detailed-results} shows detailed results. EVA-H, Hibou-B \& Hibou-L \cite{nechaev2024hibou} and CHIEF \cite{wang2024pathology} were evaluated. We evaluated H-Optimus-0 \cite{hoptimus0} on the same tasks, but obtained very poor results for this model. Hence, we decided not to report them. 
\paragraph{Common Evaluation Framework.}
All models are evaluated using a Multiple Instance Learning (MIL) \cite{ilse2018attention} framework with gated attention-based aggregation. Tile-level embeddings are first extracted from whole slide images using the frozen foundation model encoder and stored as H5 files. These embeddings are first projected to a hidden space via a fully connected layer, then processed by a gated attention module that computes tile-level importance weights in a fixed 128-dimensional attention space. The weighted aggregation of tile representations is performed in the hidden space (see Table~\ref{tab:histo_mlp_sizes} for model-specific dimensions). Training uses AdamW optimizer with constant learning rate and early stopping based on validation loss. Each task is evaluated using 5-fold cross-validation repeated across 5 random seeds.

\paragraph{Classification Tasks (Sjögren's Disease).}
Classification tasks (focus score and diagnosis) operate at the patient level, aggregating all slides from a given patient into a single prediction. Cross-validation uses stratified k-fold splitting to preserve class balance across folds. The training objective combines a bag-level cross-entropy loss with an instance-level clustering loss \cite{lu2021data} that encourages discriminative tile representations. Training uses a learning rate of $10^{-5}$ and weight decay of $0.08$. The validation metric used is the AUROC.

\paragraph{Regression Tasks (IBDs).}
Regression tasks (normalized modified Naini-Cortina and normalized modified Riley scores) operate at the slide level, where each slide receives an independent prediction. Cross-validation uses grouped k-fold splitting to ensure slides from the same patient remain together, preventing data leakage. The training objective is MSE loss. Training uses a learning rate of $1.5 \times 10^{-4}$ and weight decay of $0.5$. The validation metric used is the Pearson correlation coefficient.

\begin{table}[htbp]
\centering
\caption{MIL model dimensions for each foundation model.}
\label{tab:histo_mlp_sizes}
\begin{tabular}{lc}
\toprule
\tableheaderstyle \textbf{Model} & \textbf{Embedding dim $\rightarrow$ Hidden dim $\rightarrow$ Attention dim} \\
\midrule
EVA-H & 768 $\rightarrow$ 256 $\rightarrow$ 128 \\
Hibou-B & 768 $\rightarrow$ 256 $\rightarrow$ 128 \\
Hibou-L & 1024 $\rightarrow$ 512 $\rightarrow$ 128 \\
CHIEF & 768 $\rightarrow$ 256 $\rightarrow$ 128 \\
\bottomrule
\end{tabular}
\end{table}

\paragraph{Results.} Table~\ref{tab:histo-detailed-results} presents the results where the final prediction for each seed is obtained by averaging predictions across the 5 folds for regression tasks, or by majority vote for classification tasks. We report the average score (AUROC for classification tasks and Pearson coefficient for regression tasks) as well as the standard deviation across the 5 seeds.
\begin{table}[!ht]  
\centering
\caption{EVA-H performances on I\&I tasks. Higher is better. Bold and underline represent the first and second best model per task. Tasks are described in Section~\ref{histology-tasks-presentation}.}
\label{tab:histo-detailed-results}
\begin{tabular}{lcccc}
\toprule
\tableheaderstyle Task & Eva-H & Hibou-B & Hibou-L & CHIEF \\
\midrule
Focus score & $\underline{0.908 \pm 0.010}$ & $0.834 \pm 0.034$ & $0.856 \pm 0.021$ & $\mathbf{0.924 \pm 0.002}$ \\
Diagnosis    & $\underline{0.886 \pm 0.049}$ & $0.791 \pm 0.018$ & $0.813 \pm 0.051$ & $\mathbf{0.933 \pm 0.003}$ \\
Modified normalized Naini-Cortina score   & $0.774 \pm 0.008$ & $\underline{0.793 \pm 0.007}$ & $\mathbf{0.803 \pm 0.005}$ & $0.749 \pm 0.004$ \\
Modified normalized Riley score   & $0.902 \pm 0.004$ & $\underline{0.908 \pm 0.005}$ & $\mathbf{0.911 \pm 0.007}$ & $0.904 \pm 0.007$ \\
\bottomrule
\end{tabular}
\end{table}

\subsection{Cross-species and cross-technologies analysis} \label{apx:species-techno}

\subsubsection{Nearest neighbor rank evolution}\label{apx:nn-rank-evolution}

In this section, only input embeddings are considered. For each mouse gene having an ortholog, we computed cosine similarity between its embedding and the embeddings of all 16,168 human genes (that have mouse orthologs). We then ranked the human genes by decreasing similarity and recorded the position of the true ortholog. A rank of 1 indicates the true ortholog is the nearest neighbor (best case); higher ranks indicate more human genes are closer to the mouse gene than its true ortholog. Gene sets for each group were defined by a domain expert using NCBI database queries.

\subsubsection{Contextualized gene embedding} \label{apx:context-gene-emb} 

We randomly selected 40 samples per dataset and encoded sequences of 1000 genes. We collected contextualized gene embeddings for layer 30 (N-1), as it corresponds to the last layer before gene representations collapse into a low-dimensional structure. Visualizations in Figure~\ref{fig:panel_alignment}b correspond to gene embeddings with non-zero expression that were projected into a 2-D representation with first a PCA with 50 components, then a UMAP with 2 components and $min\_dist=0.5$.

\subsection{Masking ratio scheduling}\label{apx:masking-ratio}

Dynamic masking ratio scheduling during masked language model pretraining has shown benefits in natural language processing, including improved embeddings, faster convergence, and better downstream performance~\cite{wettig2023should,ankner2024dynamic,yoon2025currimae}. This progressive approach varies task difficulty throughout training, allowing the model to develop different representational strategies at different stages. We investigated whether similar benefits transfer to masked gene expression (MGE) pretraining in EVA-RNA.

We evaluated four masking ratio strategies on an intermediate-sized model (37.3M parameters) trained for 30,000 steps (3.5B tokens) with identical configurations otherwise: a constant 50\% masking ratio serving as our baseline; uniform random sampling between 20\% and 80\% at each step; a linearly decreasing schedule from 95\% to 50\%; and a linearly increasing schedule from 5\% to 50\%. All experiments used three datasets (ImmunAtlas bulk RNA-seq, ImmunAtlas microarray, and MurinAtlas), PCA-initialized external knowledge embeddings, warm-up-cosine learning rate schedule, and pure MGE loss.

\begin{figure}[htbp]
    \centering
    \begin{minipage}[c]{0.52\textwidth}
        \centering
        \includegraphics[width=\textwidth]{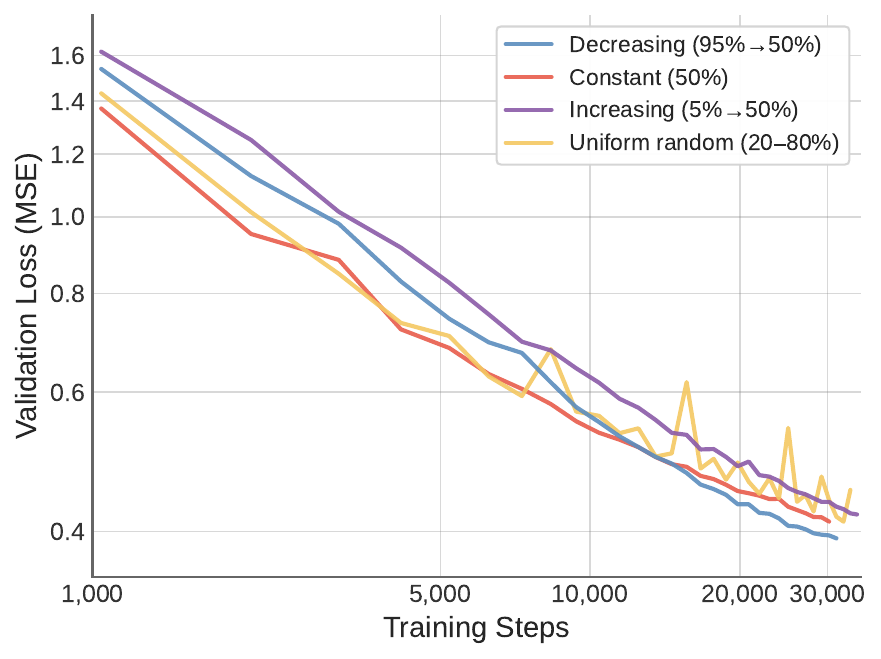}
    \end{minipage}
    \hfill
    \begin{minipage}[c]{0.44\textwidth}
        \centering
        \small
        \begin{tabular}{@{}lc@{}}
        \toprule
        \tableheaderstyle Schedule & Val loss \\
        \midrule
        Constant (50\%) & 0.411 \\
        Uniform random (20--80\%) & 0.451\textsuperscript{*} \\
        Increasing (5\%$\rightarrow$50\%) & 0.420 \\
        Decreasing (95\%$\rightarrow$50\%) & \textbf{0.392} \\
        \bottomrule
        \multicolumn{2}{l}{\footnotesize \textsuperscript{*}High training instability.}
        \end{tabular}
    \end{minipage}
    \caption{Masking ratio scheduling experiment. \textit{Left}: Validation loss trajectories over 30,000 training steps. \textit{Right}: Final validation loss for each strategy. The decreasing schedule achieves the best performance while random scheduling introduces substantial training instability.}
    \label{fig:masking-ratio}
\end{figure}

As shown in Figure~\ref{fig:masking-ratio}, the decreasing schedule achieves the lowest final validation loss despite starting with the most challenging reconstruction task at 95\% masking. We believe that high masking ratios early in training force the model to rely heavily on cross-gene contextual relationships, learning robust representations before transitioning to easier reconstruction with more available context. In contrast, the increasing schedule shows slower early convergence and a higher final loss, as the model initially receives abundant context that provides a limited learning signal. Most notably, uniform random scheduling not only yields the worst final performance but also exhibits substantial training instability with pronounced loss spikes, suggesting that abrupt changes in task difficulty between batches interfere with optimization dynamics. Based on these results, we adopted the decreasing masking ratio schedule (95\%$\rightarrow$50\%) for EVA-RNA pretraining, and rescheduled it further from 50\% to a final 15\% mid-training after a first validation loss plateau was reached.

\subsection{EVA-RNA's external knowledge details}\label{apx:prior}

\subsubsection{external knowledge embeddings computation}\label{apx:prior-compute}

This section details how external knowledge embeddings were computed for each source.
Table~\ref{tab:prior_stats} summarizes the main statistics.

\paragraph{scGPT.}
scGPT-human weights available in this \href{https://drive.google.com/drive/folders/1oWh_-ZRdhtoGQ2Fw24HP41FgLoomVo-y}{drive folder} were used. They are provided by \href{https://github.com/bowang-lab/scGPT}{the official GitHub repository}. From this model, the embedding matrix was fetched and queried. EVA-RNA's mouse genes having a human ortholog contained in scGPT vocabulary were initialized with their human ortholog embedding. This resulted in 43,089 gene embeddings of size 512, among which 15,991 (37.1\%) are mouse genes.
\paragraph{ESM2.}
Genes were mapped to their corresponding protein sequences using species-specific annotation databases. When multiple isoforms exist for a gene, the first canonical isoform is selected. The resulting sequences were tokenized and embedded using \href{https://huggingface.co/facebook/esm2_t33_650M_UR50D}{ESM2 650M model} available on Huggingface, with mean-pooling across sequence positions to generate fixed-dimension gene embeddings. This resulted in 39,363 gene embeddings of size 1280, among which 15945 (40.5\%) are mouse genes.
\paragraph{NCBI.}
NCBI (National Center for Biotechnology Information) is a U.S. government resource that maintains biomedical and genomic databases, including GenBank (DNA sequences) and PubMed (scientific literature).
We used the NCBI gene summary table available \href{https://ftp.ncbi.nlm.nih.gov/gene/DATA/}{here} (\emph{gene\_summary.gz}). Descriptions of genes of interest were embedded with \href{https://platform.openai.com/docs/models/text-embedding-3-small}{OpenAI text-embedding-3-small} model. This resulted in 43,885 gene embeddings of size 1536, among which 16,182 (36.9\%) are mouse genes.
\paragraph{UniProt.}
UniProt (Universal Protein Resource) is a comprehensive database that catalogs protein sequences and functional information. It provides detailed and curated annotations on protein structure, function, and pathways.
The following tables were downloaded: the \href{https://www.uniprot.org/uniprotkb?query=reviewed%3Atrue&facets=model_organism%3A9606}{human reviewed}, \href{https://www.uniprot.org/uniprotkb?query=*&facets=reviewed%3Afalse%2Cmodel_organism%3A9606}{human unreviewed}, \href{https://www.uniprot.org/uniprotkb?query=reviewed%3Atrue&facets=model_organism%3A10090}{mouse reviewed} and \href{https://www.uniprot.org/uniprotkb?query=*&facets=reviewed%3Afalse%2Cmodel_organism%3A10090}{mouse unreviewed}.
For each protein, NCBI gene identifiers were mapped to gene symbols, and UniProt tables were queried to extract functional annotations from the "Function [CC]" column. When multiple entries exist for the same protein, reviewed versions were prioritized over unreviewed ones, and only the first occurrence per gene symbol was retained. The extracted functional text descriptions were then embedded with \href{https://platform.openai.com/docs/models/text-embedding-3-small}{OpenAI text-embedding-3-small} model. This resulted in 31,919 gene embeddings of size 1536, among which 13,526 (42.4\%) are mouse genes.
\paragraph{KGE.}
We computed embeddings of genes using RotatE \cite{sun2019rotateknowledgegraphembedding} method on a biomedical knowlege graph. Quality of embeddings was then assessed on a range of separate classification tasks.
Similar to scGPT, mouse gene embeddings were initialized with their human ortholog embedding. This resulted in 36,399 gene embeddings of size 512, among which 16,007 (44\%) are mouse genes.

\begin{table}[H]
\centering
\caption{Summary of gene embeddings per external knowledge source}
\label{tab:prior_stats}
\begin{tabular}{llcccc}
\toprule
\tableheaderstyle \textbf{Source} & \textbf{Modality} & \textbf{Embedding Size} & \textbf{Total Genes} & \textbf{Human Genes} & \textbf{Mouse Genes} \\
\midrule
scGPT   & Single-cell RNA-seq & 512  & 43,089 & 27,098 & 15,991 \\
ESM2    & Protein structure & 1280 & 39,363 & 23,418 & 15,945 \\
NCBI    & Text (gene) & 1536 & 43,885 & 27,703 & 16,182 \\
UniProt & Text (protein) & 1536 & 31,919 & 18,393 & 13,526 \\
KGE   & Knowledge graph & 512  & 36,399 & 20,392 & 16,007\\
\bottomrule
\end{tabular}
\end{table}

\subsubsection{EVA-RNA: Embedding dimension ablation}\label{apx:emb-dimension-ablation}

We investigated the impact of reducing the dimension of external knowledge embeddings on model performance and training efficiency. For an intermediate-size model (hidden size 256, 24 layers, $\sim$ 55M parameters), we compared three settings: the default reduced size of 32, an intermediate size of 128, and no reduction (256). Results (see Figure~\ref{fig:prior-panel}c) demonstrated that the intermediate embedding size of 128 achieved both faster convergence and superior final performance compared to the highly compressed default of 32, while showing comparable training dynamics to the full-sized embeddings (256). This suggests an optimal trade-off where moderate dimensionality reduction maintains model expressiveness while improving parameter efficiency. We did not run further experiments for the 768h / 300M model and decided to use a reduced embed size of 256.

\subsubsection{EVA-RNA: external knowledge sources ablation}\label{apx:prior-ablation}
We evaluated an intermediate-sized model (256 hidden dimensions, 24 layers, $\sim$ 55M parameters) across eight configurations: no external knowledge source used, each of the five external knowledge sources individually (KGE, ESM2, UniProt, NCBI, and scGPT), all sources except ESM2, and all five sources combined. As shown in Figure~\ref{fig:prior-panel}d, initializing embeddings from all five sources yields the best performance, achieving both faster convergence and lower final validation loss. All training runs were terminated via early stopping.

\subsection{Layer-wise intrinsic dimensionality analysis}\label{apx:intrisic-dimensionality}

\begin{figure}[ht!]
    \centering
    \includegraphics[width=\textwidth]{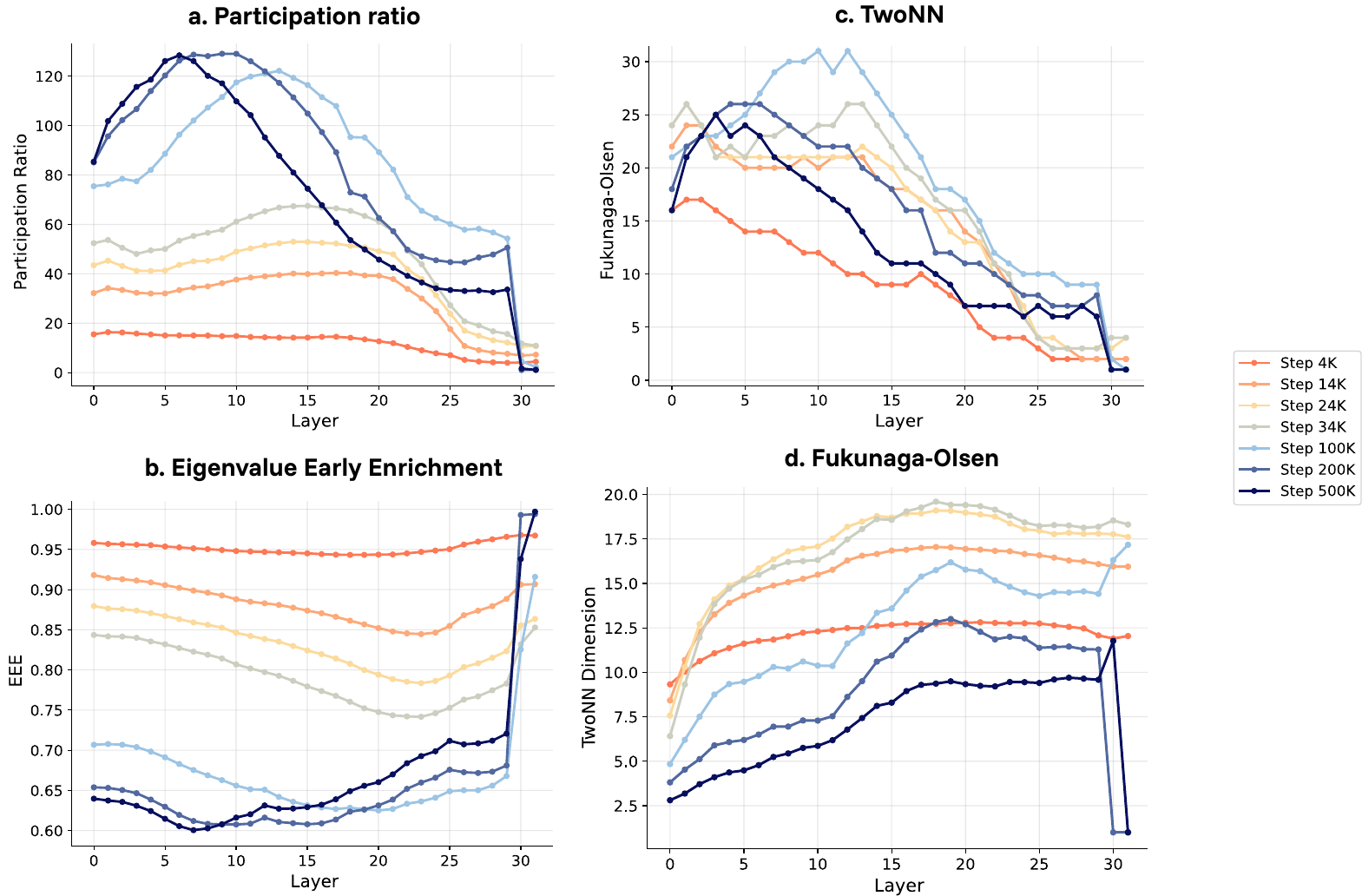}
    \caption{Intrinsic dimensionality estimation of contextualized gene embeddings across layers throughout training using four dimensionality metrics.}
    \label{fig:intrisic-dimensionality}
\end{figure}

To characterize how information is encoded within EVA-RNA representations at different depths and training stages, we estimated the intrinsic dimensionality of contextualized gene embeddings\footnote{We extracted contextualized gene embeddings from each of the 32 transformer layers (indexed from 0 to 31) by registering forward hooks during inference. For each layer, we collected embeddings from 100 samples across multiple bulk RNA-seq datasets, randomly selecting 1{,}200 genes per sample for a total of up to 50{,}000 gene embeddings per layer.} throughout the transformer layers and at various training steps using four dimensionality estimators (Figure \ref{fig:intrisic-dimensionality}). Here is a breakdown of each of them.

\subsubsection{TwoNN}

The TwoNN estimator~\cite{facco2017estimating} provides a robust, assumption-light estimate of the ID of a point cloud by exploiting the statistics of nearest-neighbor distance ratios, without requiring explicit eigenvalue decomposition or a choice of variance threshold.

Given a set of gene embeddings $\mathbf{E} = \{\mathbf{e}_1, \ldots, \mathbf{e}_n\} \subset \mathbb{R}^d$, for each point $\mathbf{e}_i$ we compute the distances $r_1^{(i)}$ and $r_2^{(i)}$ to its first and second nearest neighbors, respectively, and form the ratio $\mu_i = r_2^{(i)} / r_1^{(i)}$. Under the assumption that the data locally follows a uniform distribution on a $d_{\text{ID}}$-dimensional manifold, these ratios follow a Pareto distribution with shape parameter $d_{\text{ID}}$. The intrinsic dimensionality is then estimated via maximum likelihood:
\begin{equation}
    \hat{d}_{\text{ID}} = \left( \frac{1}{n} \sum_{i=1}^{n} \log \mu_i \right)^{-1}
\end{equation}

Intuitively, in high-ID spaces, the second neighbor tends to be only marginally farther than the first (small $\log \mu_i$), yielding a large estimate; in low-ID spaces, the second neighbor is relatively much farther, yielding a small estimate. A higher intrinsic dimensionality indicates that the representation populates more of its available degrees of freedom, suggesting richer and more diverse feature encoding.

\subsubsection{Participation ratio}
Given a matrix of gene embeddings $\mathbf{E} \in \mathbb{R}^{n \times d}$, where $n$ is the number of genes and $d$ is the hidden dimension, we computed the eigenvalues $\lambda_1, \lambda_2, \ldots, \lambda_k$ of the covariance matrix. The participation ratio is then defined as:
\begin{equation}
    \text{PR}(\mathbf{E}) = \frac{\left( \sum_{i=1}^{k} \lambda_i \right)^2}{\sum_{i=1}^{k} \lambda_i^2}
\end{equation}
The participation ratio provides a continuous, differentiable measure of the intrinsic dimensionality of a representation space. When all eigenvalues are equal (maximum spread), $\text{PR} = k$; when a single eigenvalue dominates, $\text{PR} \approx 1$.

\subsubsection{Eigenvalue Early Enrichment}
To assess the uniformity of variance distribution across embedding dimensions, we computed the Eigenvalue Early Enrichment (EEE) score~\cite{marbut2023reliable}. Given a matrix of gene embeddings $\mathbf{E} \in \mathbb{R}^{n \times d}$, where $n$ is the number of genes and $d$ is the hidden dimension, we computed the eigenvalues $\lambda_1, \lambda_2, \ldots, \lambda_d$ of the covariance matrix, sorted in decreasing order. The EEE score is then defined as:
\begin{equation}
    \text{EEE} = \frac{\text{AUC}(X_{\text{EEE}} - Y_{\text{ref}})}{\frac{1}{2} d v}
\end{equation}
where $X_{\text{EEE}}$ is the cumulative sum of eigenvalues, $Y_{\text{ref}}$ is the expected linear cumulative sum under uniform variance distribution, and $v = \sum_{i=1}^{d} \lambda_i$ is the total variance. The EEE score measures the area between the observed cumulative eigenvalue curve and the ideal linear reference, normalized by the maximum possible area. Values close to zero indicate well-spread representations where variance is distributed evenly across dimensions, while values approaching one indicate that variance is concentrated in a small number of dimensions.
\subsubsection{Fukunaga-Olsen}
The intrinsic dimensionality (ID) was also estimated using the Fukunaga-Olsen method~\cite{fukunaga1971algorithm}, which counts the number of normalized eigenvalues of the covariance matrix exceeding a threshold $T$:
\begin{equation}
    \text{ID} = \left| \left\{ i: \lambda_i > T \right\} \right|, \quad T = \frac{\max_j(\lambda_j)}{10}
\end{equation}
where $\lambda_1, \lambda_2, \ldots, \lambda_d$ are the eigenvalues sorted in decreasing order. The ID ranges from 1 (all variance concentrated in a single dimension) to $d$ (variance uniformly spread across all dimensions).

\subsection{Zero-shot perturbation benchmark details}\label{apx:zsp-benchmark}

The zero-shot perturbation (ZSP) benchmark evaluates whether pretrained representations generalize to new disease-drug combinations without task-specific fine-tuning. Table~\ref{tab:zsp-drugs} summarizes the 28 therapeutic drugs included in the benchmark, along with their molecular targets and the diseases for which they were evaluated.

\begin{table}[htbp]
\centering
\caption{Drugs and molecular targets in the ZSP benchmark. For each drug, we list the mechanism of action, the target gene(s) perturbed \textit{in silico}, and the diseases evaluated. Disease abbreviations: AD = Atopic Dermatitis, Pso = Psoriasis, HS = Hidradenitis Suppurativa, PsA = Psoriatic Arthritis, RA = Rheumatoid Arthritis, CD = Crohn Disease, UC = Ulcerative Colitis.}
\label{tab:zsp-drugs}
\begin{adjustbox}{width=\textwidth}
\begin{tabular}{llll}
\toprule
\tableheaderstyle Drug & Mechanism & Target Gene(s) & Diseases Evaluated \\
\midrule
\multicolumn{4}{l}{\textit{IL-17 pathway inhibitors}} \\
Secukinumab & Anti-IL-17A & IL17A & AD, Pso, HS, PsA, CD \\
Ixekizumab & Anti-IL-17A/F & IL17A, IL17F & Pso, PsA \\
Brodalumab & Anti-IL-17RA & IL17RA & Pso \\
\midrule
\multicolumn{4}{l}{\textit{IL-23 pathway inhibitors}} \\
Guselkumab & Anti-IL-23 & IL23A & AD, Pso, HS, PsA, CD, UC \\
\midrule
\multicolumn{4}{l}{\textit{IL-4/IL-13 pathway inhibitors}} \\
Dupilumab & Anti-IL-4R$\alpha$ & IL4R & AD, Pso \\
Tralokinumab & Anti-IL-13 & IL13 & AD, UC \\
\midrule
\multicolumn{4}{l}{\textit{Other interleukin inhibitors}} \\
Nemolizumab & Anti-IL-31RA & IL31RA & AD \\
Etokimab & Anti-IL-33 & IL33 & AD \\
Bempikibart & Anti-IL-7R & IL7R & AD \\
Bermekimab & Anti-IL-1$\alpha$ & IL1A & UC \\
\midrule
\multicolumn{4}{l}{\textit{TNF inhibitors}} \\
Etanercept & Anti-TNF & TNF & AD, Pso, HS, PsA, CD, UC \\
\midrule
\multicolumn{4}{l}{\textit{JAK inhibitors}} \\
Upadacitinib & JAK1 inhibitor & JAK1 & AD, HS, PsA, CD, UC \\
\midrule
\multicolumn{4}{l}{\textit{S1P receptor modulators}} \\
Ozanimod & S1PR1/5 modulator & S1PR1, S1PR5 & AD, CD, UC \\
Fingolimod & S1PR1/3/4/5 modulator & S1PR1, S1PR3, S1PR4, S1PR5 & AD, UC \\
\midrule
\multicolumn{4}{l}{\textit{OX40 pathway inhibitors}} \\
Rocatinlimab & Anti-OX40 & OX40 & AD \\
\midrule
\multicolumn{4}{l}{\textit{Other mechanisms}} \\
Tapinarof & AhR agonist & AHR & AD, Pso \\
Rituximab & Anti-CD20 & CD20 & UC \\
Crisaborole & PDE4 inhibitor & PDE4 & AD, Pso, PsA \\
Afimkibart & Anti-TL1A & TL1A & CD \\
Obefazimod & miR-124 inhibitor & MIR124-1 & UC \\
\bottomrule
\end{tabular}
\end{adjustbox}
\end{table}

\subsection{Contributing authors}
The listing of authors is in alphabetical order based on their last names. Names marked with an asterisk (*) indicate the work was done during an internship.
\newline
\newline
Ethan Bandasack*, Vincent Bouget, Apolline Bruley, Yannis Cattan, Charlotte Claye, Matthew Corney, Julien Duquesne, Karim El Kanbi, Aziz Fouché, Pierre Marschall, Francesco Strozzi.

\end{document}